\documentclass[11pt,a4paper]{article}
\usepackage{jcappub}

\usepackage{bm}
\usepackage{epsfig}
\usepackage{citesort}
\usepackage{graphicx}
\usepackage{amssymb}
\usepackage{color}
\usepackage{subfigure}

\begin{document}


\title{Star-forming galaxies as the origin of diffuse high-energy backgrounds: Gamma-ray and neutrino connections, and implications for starburst history}

\author[a]{Irene Tamborra,}
\author[a]{Shin'ichiro Ando}
\author[b]{and Kohta Murase}

\affiliation[a]{GRAPPA Institute, University of Amsterdam, Science Park 904, 1098 XH, Amsterdam, The Netherlands}
\affiliation[b]{Hubble Fellow -- Institute for Advanced Study, 1 Einstein Dr., Princeton, NJ 08540, USA}

\emailAdd{i.tamborra@uva.nl}
\emailAdd{s.ando@uva.nl}
\emailAdd{murase@ias.edu}

\abstract{
Star-forming galaxies have been predicted to contribute considerably to the diffuse gamma-ray background as they are guaranteed reservoirs of cosmic rays.  Assuming that the hadronic interactions responsible for high-energy gamma rays also produce high-energy neutrinos and that $\mathcal{O}(100)$~PeV cosmic rays can be produced and confined in starburst galaxies, we here discuss the possibility that star-forming galaxies are also the main sources of the high-energy neutrinos observed by the IceCube experiment.  
First, we compute the diffuse gamma-ray background from star-forming galaxies, adopting the latest {\it Herschel} PEP/HerMES luminosity function and relying on the correlation between the gamma-ray and infrared luminosities reported by {\it Fermi} observations.  
Then we derive the expected intensity of the diffuse high-energy neutrinos from star-forming galaxies including normal and starburst galaxies.  Our results indicate that starbursts, including those with active galactic nuclei and galaxy mergers, could be the main sources of the high-energy neutrinos observed by the IceCube experiment.  
We find that assuming a cosmic-ray spectral index of 2.1--2.2 for all starburst-like galaxies, our predictions can be consistent with both the {\it Fermi} and IceCube data, but larger indices readily fail to explain the observed diffuse neutrino flux.  
Taking the starburst high-energy spectral index as free parameter, and extrapolating from GeV to PeV energies, we find that the spectra  harder than $E^{-2.15}$ are likely to be excluded by the IceCube data, which can be more constraining than the {\it Fermi} data for this population.
}
\maketitle

\section{Introduction}                        \label{sec:introduction}
The existence of the extragalactic diffuse gamma-ray background (EGRB) has been revealed long ago~\cite{Fichtel1977}.
Its importance is related to the fact that it encodes unique information about high-energy processes in the Universe.  
The EGRB has an isotropic sky distribution by its definition, and is contributed by gamma rays produced
from leptonic processes or by cosmic rays interacting with the surrounding gas and radiation fields.
It is most probably made of the superposition of contributions from unresolved extragalactic sources including
star-forming galaxies, active galactic nuclei (AGN), gamma-ray bursts, intergalactic shocks produced by structure formation, etc.~(see Refs.~\cite{Dermer:2007fg,Inoue:2011bp,Collaboration:2010gqa,Ackermann:2012vca,Stecker:2010di,DiMauro:2013zfa}).  

Before {\it Fermi} and IceCube data being available, star-forming galaxies had been expected to contribute to the diffuse GeV-TeV gamma-ray background, see for example  Refs.~\cite{sto+76,sj99,Pavlidou:2002va,tho+06}, and possibly to the cumulative sub-PeV neutrino background~\cite{lw06,Thompson:2006np,Lacki:2010vs,Murase:2013rfa,kat+13}. 
Their main emission component above GeV energies is due to cosmic-ray interactions with the diffuse gas.
Diffuse emission from the Milky-Way galaxy (MW) is, in fact, the brightest feature of the gamma-ray sky.  
Gamma rays primarily arise from the decay of $\pi^0$ mesons produced in inelastic collisions of cosmic rays with the interstellar gas.  As the MW, it is expected that other star-forming galaxies emit gamma rays through the same interactions providing substantial contribution to the EGRB. 

A special subset of star-forming galaxies is a less numerous, but individually more luminous population: The starburst galaxies.  Starbursts undergo an epoch of star formation in a very localized region at an enhanced rate (i.e., $10$--$100\ M_\odot/$year) in comparison to normal galaxies such as the MW ($1$--$5\ M_\odot/$year).
This activity is often triggered by galaxy mergers or by galactic bar instabilities where the 
dynamical equilibrium of the interstellar gas is disturbed and leads to the formation of
high-density gas regions, usually at the center of the galaxy.
For example, M82 and NGC 253 are located relatively nearby ($D \sim
2.5$--4.0~Mpc) and they can be defined as prototypical starbursts; 
each of them has an intense star-forming region in the center (with a
radius of $200$~pc). These starbursts are expected to have high supernova
rates of about 0.03--0.3~yr$^{-1}$ (see Ref.~\cite{Lacki:2010vs} for a
detailed study on the characteristics of these sources). 
Observationally, EGRET detected only the MW and Large Magellanic Cloud~\cite{Abdo:2010pq}, 
but {\it Fermi} has also detected the Small Magellanic Cloud~\cite{Abdo:2010d}, the
starburst galaxies M82 and NGC 253~\cite{Abdo:2010f} as well as
starbursts with Seyferts NGC 4945 and NGC 1068~\cite{Abdo:2010g}, and some
others more recently~\cite{Ackermann:2012vca,Hayashida:2013wha}. 

Our understanding of how galaxies form and evolve has increased
dramatically over the last decade (see Ref.~\cite{Dunlop:2012pt} for a review). 
A wide range of surveys shows that the star formation rate density rises up to $z
\sim 1$--2 and reveal that the bulk of the stellar mass density seen in the
universe today was formed between $z \sim 1$ and $\sim$ 2. 
Understanding why the universe was much more active in the past and which processes or
mechanisms drove galaxy evolution is one of the most important
goals in the field of galaxy formation. 
In particular, one has to consider the energy absorbed by dust and
re-emitted at longer wavelengths, in the infrared (IR) or sub-millimeter
(sub-mm) range. Dust is responsible for obscuring the ultraviolet and
optical light from galaxies. Since star formation is affected by dusty
molecular clouds, far-IR and sub-mm data, where the absorbed radiation
is re-emitted, are very important to provide a complete picture of the
star formation history.

Surveys of dust emission performed with former satellites exploring
the Universe in the mid- and far-IR domain (IRAS, ISO, and \emph{Spitzer})
allowed the first studies of the IR-galaxy luminosity function (comoving number density 
of galaxies per unit luminosity range) up to $z
\sim 2$~\cite{Rodighiero:2009up,Dole:2006,Hauser:2001xs,Puget:1996fx}. 
The detection of large numbers of high-$z$ sources, however, was
not achievable before the \emph{Herschel Space Observatory}~\cite{Gruppioni:2013jna}, 
which has recently provided an estimate of the luminosity function up to $z\simeq4$. 
Strong evolution in both luminosity and density has been found, indicating that IR galaxies
were more luminous and more numerous in the past~\cite{Gruppioni:2013jna,Devlin:2009qn,Bethermin:2012jd,Barger:2012st,Casey:2014hya}.  
 {\it Herschel} even enabled, for the first time, the estimate of the luminosity functions of specific galaxy populations: 
normal galaxies, starbursts, and star-forming galaxies with obscured or low-luminosity AGNs,  all of them contributing to 
the star-formation rate. We note, however, that our knowledge of the far-IR luminosity function
is still affected by substantial uncertainties, because of source
confusion and low detector sensitivity~\cite{Pilbratt:2010mv}. 

The IceCube neutrino observatory at the South Pole is presently the most
sensitive instrument to uncover astrophysical neutrino sources in the TeV to PeV energy range. 
The IceCube Collaboration has recently reported evidence for extraterrestrial neutrinos, after the
observation  of three PeV neutrino cascades within three years of operation
~\cite{Aartsen:2013bka,IceCubetalk,Aartsen:2014gkd}. 
The recent data set consisting of $37$ events corresponds to a spectral excess with respect to the
atmospheric background with a significance of more than $5 \sigma$~\cite{Aartsen:2014gkd}. 
This spectral excess is consistent with a diffuse neutrino flux with a $E^{-2}$ spectrum, possible break/cutoff at a few PeV, 
and isotropic sky distribution.  
The flux for each neutrino species is well described with the following
fit based on the data between $30$~TeV and $2$~PeV:
\begin{equation}
\label{fitnu}
E_{\nu}^2\ I_\nu(E_\nu) \simeq (0.95 \pm 0.3) \times 10^{-8}\ \mathrm{GeV}\ \mathrm{cm}^{-2}\ \mathrm{s}^{-1}\ \mathrm{sr}^{-1}\,
\end{equation}
within $1\sigma$ error bars. 

The energy distribution of the 37 neutrino events should provide additional clues about the candidate sources and
several explanations have been indeed proposed to explain such high-energy events (see Refs.~\cite{wax13,Anchordoqui:2013dnh} for recent reviews).  
Standard fluxes of the prompt neutrino background from the decay of charmed mesons are too low to have a significant
contribution at PeV energies~\cite{Enberg:2008te,Gaisser:2013ira}.  
A connection to cosmogenic neutrinos produced via the extragalactic
background light seems unlikely~\cite{Laha:2013lka,Roulet:2012rv},
unless one assumes the optimistic extragalactic background disfavored by
{\it Fermi} observations of gamma-ray bursts along with relatively low
maximum proton energies~\cite{Kalashev:2013vba}. 
Early studies suggested a possible connection to the Glashow resonance~\cite{Bhattacharya:2011qu}, 
but this scenario was later disfavored by a follow-up analysis~\cite{Aartsen:2013bka}. 
More exotic models such as the PeV dark matter decay scenarios have  been 
suggested too~\cite{Feldstein:2013kka,Esmaili:2013gha}. 
Galactic neutrino sources have been discussed, pointing out a possible
association with the extended region around the Galactic center (see Refs.~\cite{Ahlers:2013xia,raz13} and references therein)
or with the unidentified TeV gamma-ray sources~\cite{Fox:2013oza}. 
The Galactic halo has been discussed as one of the possible options~\cite{Ahlers:2013xia,Joshi:2013aua,Taylor:2014hya}.
However, most of the observed events do not originate from the extended region around the Galactic center which makes the scenario of the Galactic origin slightly disfavored, and GeV-PeV gamma-ray observations have already put meaningful constraints on various Galactic possibilities including the Galactic halo scenario~\cite{Ahlers:2013xia,Murase:2013rfa}. 
The isotropic diffuse flux is most naturally explained with extragalactic sources.   
Among extragalactic scenarios, various PeV neutrino sources including gamma-ray bursts, peculiar supernovae, newborn pulsars, AGN, star-forming galaxies and intergalactic shocks had already been suggested before the discovery of the IceCube spectral excess. 
The observation can be most probably associated with various extragalactic $\sim100$~PeV cosmic-ray accelerators, 
e.g. low-power gamma-ray burst jets~\cite{Murase:2013ffa}, AGN~\cite{:2013fxa,Murase:2014foa}, star-forming
galaxies including starbursts, galaxy mergers and AGN~\cite{lw06,Thompson:2006np,Lacki:2010vs, Murase:2013rfa,he+13,liu+14,kat+13,Kashiyama:2014rza,Anchordoqui:2014yva,Chang:2014hua}, intergalactic shocks and active galaxies embedded in structured regions~\cite{Murase:2013rfa}. 
Mixed scenarios of Galactic and extragalactic neutrino sources have also been discussed~\cite{Ahlers:2013xia,raz13,Fox:2013oza,Joshi:2013aua,Murase:2014foa,Padovani:2014bha}.

Since the star-forming galaxies are considered to be one of the main source classes of the diffuse gamma-ray background, we here consider
them as a potential source population of the high-energy neutrinos observed by IceCube and compare our estimations with the observed
IceCube flux. Making such a multi-messenger connection is one of the keys for revealing the origin of the diffuse neutrino flux observed by IceCube.  
As shown in Ref.~\cite{Murase:2013rfa}, the multi-messenger connection based on the measured neutrino and gamma-ray data can provide crucial ways to constrain various extragalactic scenarios such as star-forming galaxies.

We rely our computations on the latest IR luminosity function provided by {\it Herschel} PEP/HerMES up to $z \simeq 4$~\cite{Gruppioni:2013jna} as
well as the well calibrated scaling relation between gamma-ray luminosity and IR luminosity~\cite{Ackermann:2012vca}.
Compared with previous studies on the EGRB estimates due to star-forming galaxies~\cite{Pavlidou:2002va,tho+06,Fields:2010bw,Makiya:2010zt,Lacki:2012si}, 
we adopt the empirical relation obtained in the {\it Fermi} era as well as the luminosity functions of different populations of galaxies (normal galaxies, starbursts, galaxies containing obscured/low-luminosity AGN), separately, which became possible thanks to {\it Herschel}'s great statistics of various populations covering wide range of the IR spectrum.
For our \emph{canonical} model, where we assume that the spectral index of both gamma rays and neutrinos is $2.2$ from starburst-like sources, we show that 
a significant fraction ($\sim$20--50\%) of the IceCube flux may be explained with this source population given that the extrapolation from GeV to PeV energies is realized.  If the starburst population features harder spectra than $E^{-2.15}$, then the IceCube data yield more stringent constraints on
such a scenario than the {\it Fermi} data, allowing us to exclude these models.

This paper is organised  as follows.
We first calculate the EGRB intensity based on the {\it Herschel} PEP/HerMES luminosity function,
and compare the EGRB estimation with the {\it Fermi} data in Sec.~\ref{sec:IRluminosityfunction}.  
We then present the associated diffuse neutrino background in comparison with the
IceCube data in Sec.~\ref{sec:gammanu}.  
We constrain the injection spectral index of starburst-like galaxies as well as their abundance in Sec.~\ref{sec:constraints}. 
In Sec.~\ref{sec:issues}, we discuss caveats and issues of the star-forming galaxy scenario for the observed high-energy neutrinos, which should be addressed in the future.  We finally give summary and discuss perspectives in Sec.~\ref{sec:conclusions}.

\section{Extragalactic gamma-ray diffuse background from star-forming galaxies}
\label{sec:IRluminosityfunction}
In this Section, we compute the EGRB adopting the {\it Herschel} PEP/HerMES IR
luminosity function up to $z \simeq 4$~\cite{Gruppioni:2013jna} and 
the relation connecting the IR luminosity to the gamma-ray luminosity
presented by the {\it Fermi} Collaboration~\cite{Ackermann:2012vca}.
The luminosity function provides a fundamental tool to probe the
distribution of galaxies over cosmological time, since it allows us to
access the statistical nature of galaxy formation and evolution. It is
computed at different redshifts and constitutes the most direct method
for exploring the evolution of a galaxy population, describing the
relative number of sources of different luminosities counted in
representative volumes of the universe. Moreover, if it is computed for
different samples of galaxies, it can provide a crucial comparison
between the distribution of galaxies at different redshifts, in
different environments or selected at different wavelengths.

\subsection{Gamma-ray luminosity function inferred from infrared data}
\label{sec:GammaLF}
Up to now, the total emissivity of IR galaxies at high redshifts had
been poorly known because of the scarcity of \emph{Spitzer} galaxies at $z
> 2$, the large spectral extrapolations to derive the total
IR luminosity from the mid-IR, and the incomplete information on the
redshift distribution of mid-IR sources~\cite{Rodighiero:2009up}. 
{\it Herschel} is the first telescope that allows to detect the far-IR population to high
redshifts ($z < 4$--5) and to derive its rate of evolution~\cite{Gruppioni:2013jna}. 
From stellar mass function studies, one finds a clear increase of the relative fraction of
massive star-forming objects (with mass $M > 10^{11} M_\odot$) as a
function of redshifts, starting to contribute significantly to the
massive end of the mass functions at $z >
1$~\cite{Devlin:2009qn,Barger:2012st,Bethermin:2012jd,Gruppioni:2013jna,Casey:2014hya}.

The IR population does not evolve all together as a whole, but it is
composed of different galaxy classes evolving differently and independently.
According to Ref.~\cite{Gruppioni:2013jna}, among {\it detected} star-forming galaxies, 
IR observations report 38\% of normal spiral galaxies (NG), the 7\% of starbursts (SB), and the
remaining goes in star-forming galaxies containing AGN (SF-AGN) at 160\,$\mu$m,  being the latter ones star-forming
galaxies containing either an obscured or a low-luminosity  AGN but still contributing to the cosmic star-formation rate.
(Note that, in general, the {\it intrinsic} fractions are functions of redshifts.)
Here we consider all of them as gamma-ray (and neutrino) sources.
The galaxy classification has been done with the IR
spectral data, where those that have the far-IR excess with significant
ultraviolet extinction are  classified as SB, and those who exhibit mid-IR excess can be
attributed to the galaxies containing obscured or low-luminosity AGNs (SF-AGN).  Both systems (SB and SF-AGN)
 have bolometric emission dominated by star formation. The latter family, SF-AGN, according to Ref.~\cite{Gruppioni:2013jna} includes Seyferts, LINERs
and ULRIGs containing AGN; we remark as the SF-AGN family, although containing AGNs, is dominated by star-formation and not by AGN processes.
The specific star-formation rate obtained for the SB population is shown
to be on average  a factor 0.6 higher than that for the NG
population~\cite{Gruppioni:2013jna}.\footnote{Some
literature adopts the values of the specific star-formation rate in
order to define NG and SB populations. But we note that both conventions
are consistent with each other as shown in Fig.~15 of
Ref.~\cite{Gruppioni:2013jna}.}

The gamma-ray intensity is calculated with the gamma-ray luminosity function as
\begin{equation}
\label{eq:intensitygamma}
I(E_\gamma) = \int_0^{z_{\mathrm{max}}} dz \int_{L_{\gamma,
\mathrm{min}}}^{L_{\gamma, \mathrm{max}}} \frac{dL_\gamma}
{\ln(10) L_\gamma} \frac{d^2 V}{d\Omega
dz}  \ \sum_X \Phi_{\gamma,X}(L_\gamma, z) \frac{dF_{\gamma,X}(L_\gamma,
(1+z)  E_\gamma, z)}{dE_\gamma}\ e^{-\tau(E_\gamma, z)}\ ,
\end{equation}
where $\Phi_{\gamma, X}(L_\gamma, z) = d^2 N_X/dV d\log L_\gamma$ is the
gamma-ray luminosity function for each galaxy family $X = \{{\rm NG}$,
SB, SF-AGN\}, $dF_{\gamma, X}(L_\gamma, (1+z) E_\gamma, z)/dE_\gamma$ is the differential
gamma-ray flux at energy $E_\gamma$ from a source $X$ at the redshift
$z$, $d^2 V/d\Omega dz$ the comoving volume per unit solid angle per
unit redshift range. The factor $e^{-\tau(E_\gamma, z)}$ takes into account the
attenuation of high-energy gamma rays by pair production with
ultraviolet, optical and IR extragalactic background light (EBL
attenuation), $\tau(E_\gamma, z)$ being the optical depth.
In our numerical calculations we assume $z_{\mathrm{max}} \simeq 5$ and 
adopt the EBL model in Ref.~\cite{Gilmore:2011ks}.

For each population $X$, we adopt a parametric estimate of the
luminosity function  for the IR luminosity between 8
and 1000\,$\mu$m at different redshifts~\cite{Gruppioni:2013jna}:
\begin{eqnarray}
\label{eq:Herschelfit}
\Phi_{{\rm IR},X} (L_{\mathrm{IR}}, z) d\log L_{\mathrm{IR}} &=& \Phi_{{\rm
IR}, X}^{\star}(z) 
\left(\frac{L_{\mathrm{IR}}}{L_{{\rm IR},X}^\star (z)}\right)^{1-\alpha_{{\rm
IR},X}}
 \nonumber\\&&{}\times
 \exp{\left[-\frac{1}{2 \sigma_{{\rm IR},X}^2} \log^2
       \left(1+\frac{L_{\mathrm{IR}}}{L_{{\rm
	IR},X}^\star(z)}\right)\right]} d\log L_{\mathrm{IR}}\ ,
\end{eqnarray}
which behaves as a power law for $L_{\mathrm IR} \ll L_{{\rm IR},X}^\star$
and as a Gaussian in $\log L_{\mathrm{IR}}$ 
for $L_{\mathrm{IR}} \gg L_{{\rm IR},X}^{\star}$.
 We adopt the redshift evolution of $L_{{\rm
IR},X}^{\star}(z)$ and $\Phi_{{\rm IR},X}^{\star}(z)$ for each population
as in Table~8 of Ref.~\cite{Gruppioni:2013jna}, as well as
the values of $\alpha_{{\rm IR},X}$ and $\sigma_{{\rm
IR},X}$.
We then fix the normalization by fitting the data for $L_{{\rm
IR},X}^{\star}(z)$ and $\Phi_{{\rm IR},X}^{\star}(z)$ from Fig.~11 of
Ref.~\cite{Gruppioni:2013jna}.
Table~\ref{table:fit} shows the local ($z=0$) values of these parameters
as the result of such a fitting procedure.\footnote{We note
that adopting for these parameters the values directly from Table~8 of
Ref.~\cite{Gruppioni:2013jna} results in overestimate of the gamma-ray
intensity, since the table summarizes the values of $L^\star_{{\rm IR},
X}$ and $\Phi^\star_{{\rm IR}, X}$ for $0 < z < 0.3$, which are
different from the values at $z = 0$ shown in Table~\ref{table:fit}.}

By integrating the luminosity function, one obtains the IR
luminosity density $Q_{\rm IR}(z)$:
\begin{equation}
Q_{\rm IR}(z) = \int d\log L_{\rm IR} L_{\rm IR}
  \sum_{X} \Phi_{{\rm IR}, X}(L_{\rm IR}, z)\ ,
\end{equation}
which is believed to be well correlated to the cosmic star-formation
rate density.
The adopted fitting functions for the luminosity function of each family $X$ 
reproduce the total IR luminosity density data summarized in Fig.~17 of 
Ref.~\cite{Gruppioni:2013jna}.

\begin{table}
 \begin{center}
  \caption{Local values of the characteristic luminosity
  ($L^\star_{{\rm IR}, X}$) and density ($\Phi^\star_{{\rm IR}, X}$) for
  each population $X$.}
  \label{table:fit}
  \begin{tabular}{rrr}\hline\hline
   $X$ & $L^\star_{{\rm IR}, X}(z=0)$ & $\Phi^\star_{{\rm
   IR}, X}(z=0)$ \\ \hline 
   NG & $10^{9.45} L_\odot$ & $10^{-1.95}$\,Mpc$^{-3}$\\
   SB & $10^{11.0} L_\odot$ & $10^{-4.59}$\,Mpc$^{-3}$\\
   SF-AGN & $10^{10.6} L_\odot$ & $10^{-3.00}$\,Mpc$^{-3}$\\ \hline\hline
  \end{tabular}
 \end{center}
\end{table}

The {\it Fermi} data show a correlation between gamma-ray luminosity
(0.1--100~GeV) and IR luminosity (8--1000~$\mu$m).  
Although such correlation is not conclusive at present due to the limited statistics of starbursts found in gamma rays, we adopt the following scaling relation:
\begin{equation}
\label{eq:IRgamma_lum}
\log \left(\frac{L_{\gamma}}{\mathrm{erg\ s}^{-1}}\right) = \alpha \log \left(\frac{L_{\mathrm{IR}}}{10^{10} L_\odot}\right) + \beta\ ,
\end{equation}  
with $L_\odot$ the solar luminosity, $\alpha = 1.17 \pm 0.07$ and $\beta
= 39.28 \pm 0.08$~\cite{Ackermann:2012vca}.
While this parameterization is calibrated in a local volume, we assume
that it is also valid at higher redshifts (up to $z_{\mathrm{max}}
\simeq 5$) and for $10^8 L_\odot \le L_{\mathrm{IR}} \le 10^{14}
L_\odot$. 
Through this correlation relation, the gamma-ray and IR luminosity
functions are related via
\begin{equation}
 \Phi_{\gamma, X}(L_\gamma, z)\ d\log L_\gamma = \Phi_{{\rm IR}, X}
  (L_{\rm IR}, z)\ d\log L_{\rm IR}\ .
\end{equation}

The physical origin of the above correlation is qualitatively understood as follows.  Total IR luminosity at 8--1000~$\mu$m is one of the well-established tracers of the star-formation history for late-type galaxies~\cite{Kennicutt:1998zb}, since stellar emission is reprocessed by dust.  Gamma rays are produced by cosmic rays and the cosmic-ray injection rate is expected to be proportional to the star-formation rate, so the gamma-ray luminosity is expected to be correlated with the IR luminosity~\cite{Ackermann:2012vca}.  However, further studies are needed to see if such a correlation largely holds from normal to starburst galaxies.

\subsection{Gamma-ray flux}                        
\label{sec:gammaflux}
As for the distribution of gamma rays as a function of energy, we
here adopt a broken power-law fit and parameterize it as~\cite{Ando:2009nk}
\begin{eqnarray}
\label{LgammaMW}
 \frac{dF_{\gamma,X} (L_\gamma, E_\gamma, z)}{dE_\gamma} =
 a_{X}(L_\gamma, z) \times \left\{ \begin{array}{lr}
 \left(\frac{E_\gamma}{600\ {\mathrm{MeV}}}\right)^{-1.5}\ \ \
  \mathrm{s}^{-1}\ \mathrm{MeV}^{-1}\ \mathrm{for}\ E_\gamma <
  600~\mathrm{MeV} \\ 
 \left(\frac{E_\gamma}{600\ {\mathrm{MeV}}}\right)^{-\Gamma_{X}}\
  \  \ \mathrm{s}^{-1}\ \mathrm{MeV}^{-1}\ \mathrm{for}\ E_\gamma \ge
  600~\mathrm{MeV}\ .
\end{array}\right.
\end{eqnarray} 
We set the normalization $a_{X}(L_\gamma, z)$ by requiring
\begin{equation}
\int_{0.1~{\rm GeV}}^{100~{\rm GeV}} dE_\gamma\  E_\gamma
 \frac{dF_{\gamma, X}(L_\gamma,  (1+z)E_\gamma, z)}{dE_\gamma} =
 \frac{L_\gamma}{4\pi d_L^2(z)}\ ,
\end{equation}
such that the energy flux integrated over energy gives the
luminosity above 100~MeV divided by luminosity distance $d_L$ squared. The 
injection spectral index of the low energy part of the spectrum ($\Gamma
= 1.5$) has been chosen as a good fit to the gamma-ray
spectrum~\cite{Pavlidou:2002va}. 
As for the high-energy part of the spectrum, in our \emph{canonical} model, we adopt 
$\Gamma_\mathrm{NG}=2.7$, $\Gamma_\mathrm{SB}=2.2$.

Concerning the SF-AGN spectral index, within our \emph{canonical} model, we choose to adopt two different spectral indices following the evolution with the redshift of the IR luminosity density  
as in Fig.~18 and Tab.~9 of~\cite{Gruppioni:2013jna}. In fact, we subdivide the SF-AGN family in ``SF-AGN (SB),'' that resembles starburst galaxies, 
and ``SF-AGN (non-SB),''  that includes galaxies similar to the normal ones. We report in Table~\ref{tab:SFAGN} 
the fraction of SF-AGN (SB), for which we assume a SB-like spectral index, and the remaining fraction, SF-AGN (non-SB), for which we take a spectral
index equal to the one of normal galaxies, for each redshift range.
In our \emph{canonical} model, we assume the distribution reported in Table~\ref{tab:SFAGN}, 
and take: $\Gamma_\mathrm{SF-AGN(non-SB)} = \Gamma_\mathrm{NG}$ 
and $\Gamma_{\mathrm{SF-AGN(SB)}} = \Gamma_\mathrm{SB}$. 
 \footnote{In the rest of the paper, we will always distinguish among three families of star-forming galaxies referring to the three correspondent luminosity functions defined in Sec.~\ref{sec:GammaLF} for NG, SB and SF-AGN. However, in terms of the energy spectra, we assume that SF-AGN behave as normal galaxies or starbursts as a function of the redshift as pointed out in Table~\ref{tab:SFAGN}.  In fact the SF-AGN activity is dominated by star formation and not by the AGN presence.} 
Note that, in the picture of cosmic-ray diffusion, the spectral index difference 
$\delta_{\rm CR}\equiv\Gamma_\mathrm{NG}-\Gamma_\mathrm{SB}\sim0.5$ 
is interpreted as the energy dependence of the diffusion coefficient. 
\begin{table*}
 \caption{Fractions of SF-AGN (non-SB) and SF-AGN (SB) as  functions of the redshift 
 adopted in our {\it canonical} model, on the basis of Tab.~9 of Ref.~\cite{Gruppioni:2013jna}.  We assume 
 $\Gamma_\mathrm{SF-AGN(non-SB)} = \Gamma_\mathrm{NG}$ for SF-AGN (non-SB) and $\Gamma_{\mathrm{SF-AGN(SB)}} = \Gamma_\mathrm{SB}$
  for  SF-AGN (SB).}
 \center
\begin{tabular}{ccc}
\hline \hline
 redshift range &  SF-AGN (non-SB) & SF-AGN (SB)   \\ 
        &     &         \\   \hline 
 0.0$<$$z$$<$0.3 &   85 $\%$ &    15 $\%$\\
0.3$<$$z$$<$0.45 &  91 $\%$  &   9 $\%$\\
 0.45$<$$z$$<$0.6 &  99 $\%$  &  1 $\%$\\
 0.6$<$$z$$<$0.8 & 87 $\%$   & 13 $\%$\\
 0.8$<$$z$$<$1.0 &  73 $\%$  &  27 $\%$\\
 1.0$<$$z$$<$1.2 & 32 $\%$  & 68 $\%$\\
 1.2$<$$z$$<$1.7 & 75 $\%$ & 25 $\%$\\
 1.7$<$$z$$<$2.0 & 75 $\%$ & 25 $\%$\\
  2.0$<$$z$$<$2.5 &  19 $\%$ & 81 $\%$\\
 2.5$<$$z$$<$3.0 & 24 $\%$     &      76 $\%$\\
 3.0$<$$z$$<$4.2 & 28 $\%$ & 72 $\%$\\
   \hline \hline
\end{tabular}
\label{tab:SFAGN}
\end{table*}

Other emission components such as the inverse-Compton scattering due to primary electrons can 
be also relevant especially at lower energies~\cite{Lacki:2012si}.  Although we do not take them into account in this work, 
including them will not change our main results and conclusions.

\subsection{Results: Extragalactic diffuse gamma-ray background}        
 \label{sec:gammaLF}
\begin{figure}[t]
\centering
\includegraphics[angle=0, width=0.9\textwidth]{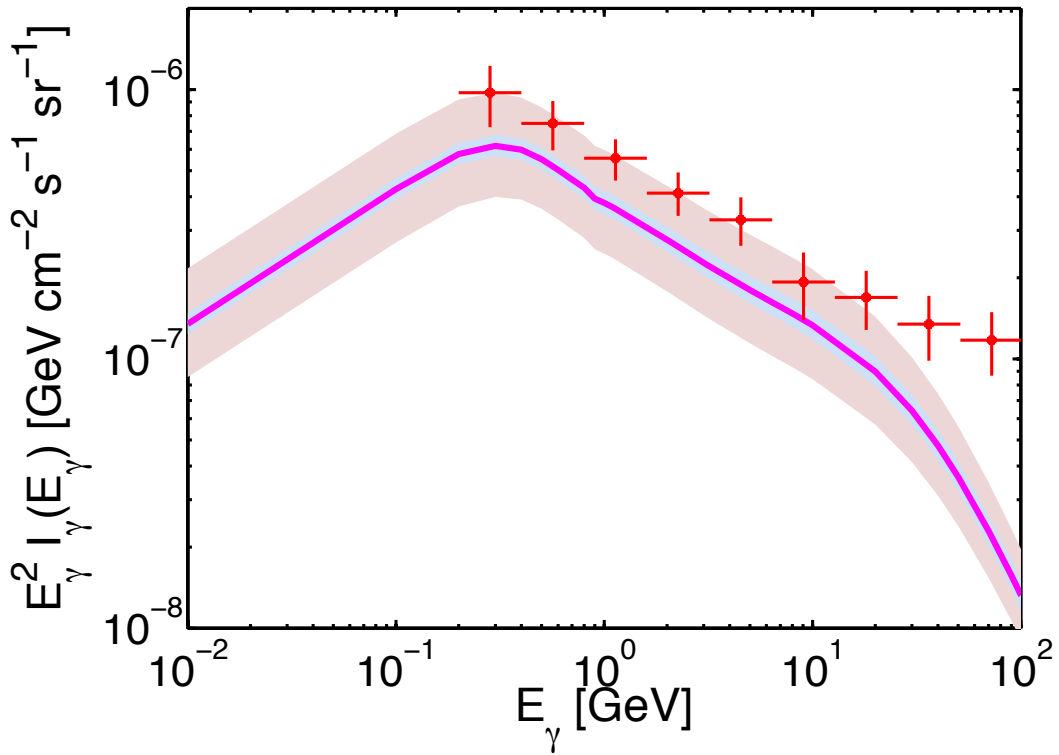}
\caption{{Diffuse gamma-ray intensity $E_\gamma^2 I_\gamma(E_\gamma)$ as a function of the energy evaluated with the {\it Herschel} PEP/HerMES IR luminosity function~\cite{Gruppioni:2013jna}. 
The magenta curve is the diffuse gamma-ray intensity obtained adopting the IR luminosity function, the light-blue band defines the error band obtained including the uncertainties in the measurements of the parameters of Eq.~(\ref{eq:Herschelfit}), while the pink band defines the uncertainties from Eq.~(\ref{eq:IRgamma_lum}). } The {\it Fermi} data~\cite{Ackermann:2012vca} are plotted in red.  
The EBL attenuation is taken into account, where the model of Ref.~\cite{Gilmore:2011ks} is used. 
 \label{fig:gammarayIR}}
\end{figure}

Adopting Eqs.~(\ref{eq:intensitygamma}), (\ref{eq:Herschelfit}) and
(\ref{eq:IRgamma_lum}), we calculate the intensity of EGRB.
Figure~\ref{fig:gammarayIR} shows $E_\gamma^2 I_\gamma(E_\gamma)$ as a
function of the energy  for our {\it canonical} model.
The magenta line defines the EGRB obtained considering the diffuse
gamma-ray intensity evaluated with the IR luminosity function, the
light-blue band defines the error band obtained including the
uncertainties in the measurements of the parameters in
Eq.~(\ref{eq:Herschelfit}), while the pink band defines the
uncertainties from Eq.~(\ref{eq:IRgamma_lum}).  The {\it Fermi}
data~\cite{Ackermann:2012vca} are plotted in red.

Our canonical EGRB estimation falls below the {\it Fermi} data, especially at $\lesssim0.3$~GeV and $\gtrsim30$~GeV energies, 
although they are very close to the upper bound of the astrophysical uncertainty region marked by the pink band.
The attenuation of the spectrum due to the EBL correction affects the high-energy part of the diffuse EGRB 
($E_{\gamma} \gtrsim30$~GeV), suggesting other sources such as BL Lac objects may contribute to match 
the observed {\it Fermi} data~\cite{Murase:2012df}.
Note that our EGRB estimation is in agreement with the estimation presented in Ref.~\cite{Stecker:2010di} adopting the IR luminosity function 
(see Ref.~\cite{Lacki:2012si} and Fig.~7 of Ref.~\cite{Ackermann:2012vca} and references therein for comparison with previous models).
 Our \emph{canonical} model explains about half of the diffuse EGRB in the 0.3--30~GeV range, which is consistent with the results presented in Ref.~\cite{Lacki:2012si}.  Considering the luminosity functions of normal and starburst galaxies only (without SF-AGN), about $\sim20$\% of the diffuse EGRB can be explained, which is also in agreement with the results of Ref.~\cite{Ackermann:2012vca}.

As for NG, we assumed that they have steeper spectra than SB (i.e., $\Gamma_\mathrm{NG} = 2.7$), 
but we note that this may not be the case at high redshifts since the column density may be higher and, 
as a result, the spectra might be closer to that of SB.
Such an effect might modify our estimation enhancing the EGRB even further at high energies. 

The EBL is the background of optical/IR light emitted by stars and AGN, including 
reprocessed components made by dust, over the lifetime of the Universe.  
Today this photon background consists of light emitted at all epochs, modified by dilution due
to the expansion of the Universe and by the redshift correction. High-energy gamma rays interact with
the EBL, and produce electron-positron pairs.
This process therefore alters the observed spectra of extragalactic high-energy sources.
We estimate that in Fig.~\ref{fig:gammarayIR}, the EBL attenuation starts at about $20$~GeV and the suppression at $100$~GeV is by a factor
of $5$. This is slightly different from Fig.~7 of Ref.~\cite{Ackermann:2012vca}, where the suppression factor at 100~GeV is less than $\sim2$.  
This is partially because we consider contributions from sources at $z>2.5$.  If we only consider sources at $z<2$, the suppression factor is reduced to $\sim3$ for the EBL model adopted here.  The remaining discrepancy of $\sim2$ can be explained by the fact that the EBL model~\cite{Franceschini:2008tp} used in Ref.~\cite{Ackermann:2012vca} is more conservative (see Refs.~\cite{Finke:2009xi,Inoue:2012bk} for comparison between the EBL model of Ref.~\cite{Franceschini:2008tp} adopted in Ref.~\cite{Ackermann:2012vca} and our model~\cite{Gilmore:2011ks}).  

The latest {\it Fermi} data show that the EGRB spectrum extends to $\sim600$~GeV~\cite{Ackermann:FS}.  
Once the EBL attenuation is taken into account, therefore, a simple power-law injection spectrum 
with {\it positive} redshift evolution cannot explain the diffuse EGRB, leading to a very-high-energy gamma-ray excess 
compared to the EBL-attenuated spectrum~\cite{Murase:2012df}.  
The very-high-energy gamma-ray excess appears because the sum of  components with NG and SB-like energy spectra lead to 
an effective spectral index $\Gamma_{\rm SF}\sim2.4$ (see Fig.~\ref{fig:gammarayIR}).   
In principle, if we consider a harder injection spectrum, e.g., only the component with SB-like energy spectrum, it is 
possible to fit the diffuse EGRB spectrum within EBL uncertainties at high redshifts (see Ref.~\cite{Murase:2013rfa}). 
However, in the star-forming galaxy scenario, this requires optimistic assumptions. 
We need a higher starburst fraction and/or smaller values of $\Gamma_{\rm SB}$ to achieve smaller effective 
spectral indices, and the sufficiently large ratio of $L_\gamma$ to $L_{\rm IR}$ is required to adjust the overall normalization.    
Although we have not included electromagnetic cascades that can relax this tension, 
our results will not change much as long as we consider $\Gamma_{\rm SB}>2$.   
Hence, although we do not exclude the possibility that the diffuse EGRB above 100~GeV energies is
explained by galaxies with a SB-like energy spectrum, 
our canonical model predicts the necessity of 
another source population at $\gtrsim100$~GeV energies, such as BL Lac objects.

\subsection{Composition of the extragalactic diffuse gamma-ray background from star-forming galaxies} 
\label{sec:highz}
\begin{figure}[b]
\centering
\includegraphics[angle=0, width=0.9\textwidth]{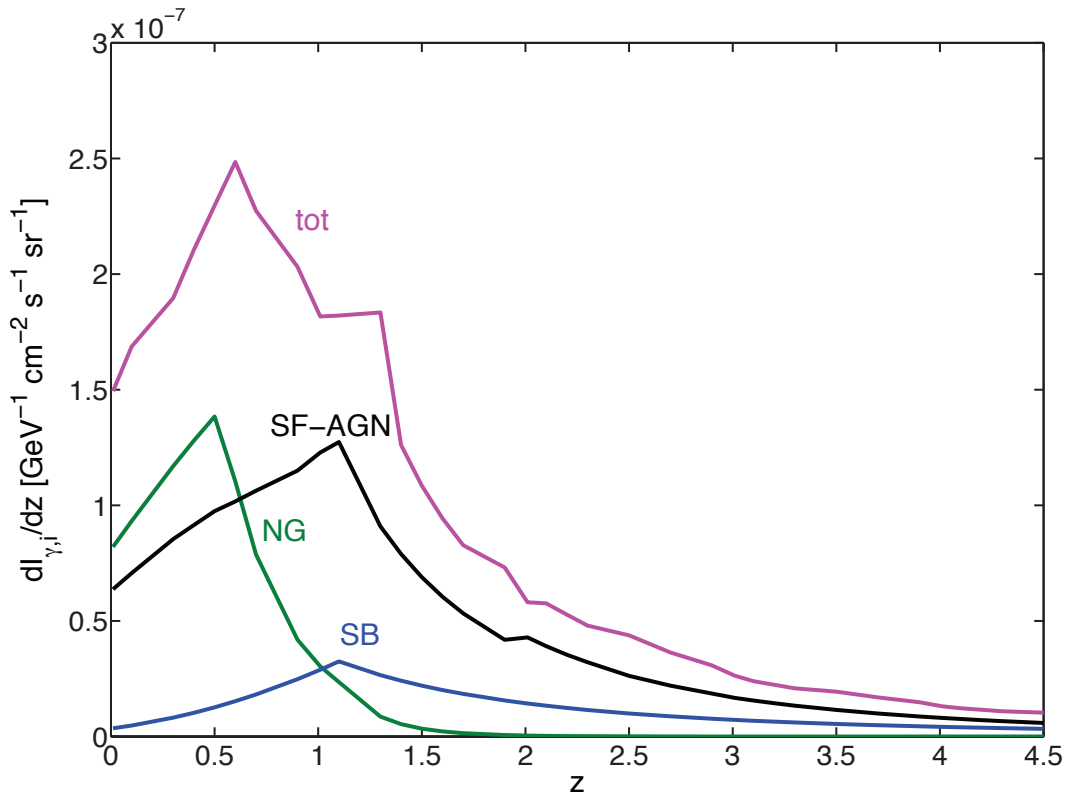}
 \caption{Differential contributions to the EGRB intensity, $d I_{\gamma,\mathrm{tot}}/dz$ at $E_\gamma = 1$~GeV: 
 NG (in green), SB (in blue), SF-AGN $=$ SF-AGN (SB) + SF-AGN (non-SB) (in black), and their sum (in magenta). Normal galaxies
 are the leading contribution up to $z \simeq 0.5$, while SF-AGN and SB dominate at higher redshifts.
 \label{fig:diff_contrib_z}}
\end{figure}
In Ref.~\cite{Gruppioni:2013jna}, the fraction of {\it detected} SB is
found to be about $7\%$ of the total galaxies and the fraction of SF-AGN
is about $48\%$ at 160\,$\mu$m.  
However, the starburst fraction is highly uncertain; it changes with the redshift,
and several recent studies speculate that it tends to increase at higher
redshifts~\cite{Gruppioni:2013jna,Barger:2012st,Casey:2014hya}.  
Moreover, some of the  star-forming galaxies containing AGN might have
SB-like  energy spectra, as we assume in our {\it canonical} model. 

The contributions to the total EGRB from different redshift ranges is shown in Fig.~\ref{fig:diff_contrib_z}, where we plot
the differential intensities as a function of $z$ for NG (in green), SB (in blue), SF-AGN ($=$ SF-AGN (SB) + SF-AGN (non-SB), in black), and their sum (in magenta) at $E_\gamma = 1$~GeV.  
The largest contribution to the EGRB comes from $z < 3$ and from SF-AGN (including SF-AGN (SB) and SF-AGN (non-SB)).  
At $z < 0.5$ the EGRB is dominated by NG ($60\%$) and SF-AGN ($40\%$).  
The abundance of normal galaxies steeply decreases for $z \ge 0.5$ and they are practically negligible at $z \ge 1.5$, while SF-AGN become the 
$80\%$ of the total EGRB and SB contribute to the remaining $20\%$.  In agreement with previous estimations, the SB contribution increases with the 
redshift, becoming $\sim$20--30\% of the total at $z >2$.

\begin{figure}[t]
\centering
\includegraphics[angle=0, width=0.9\textwidth]{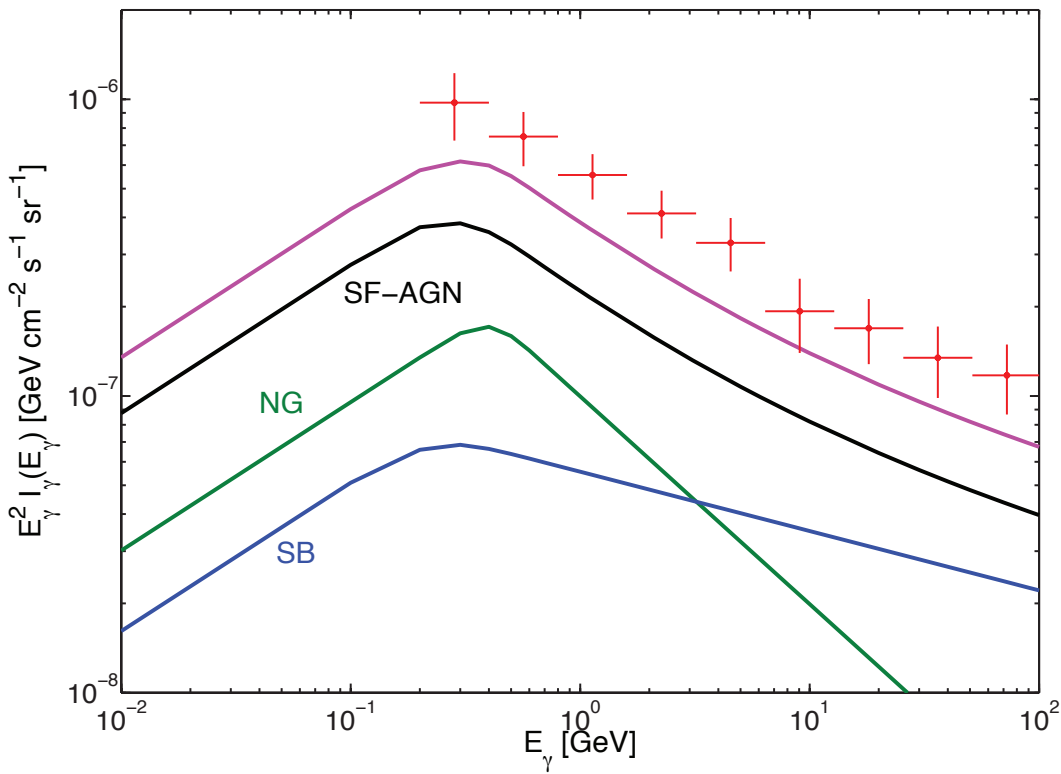}
\caption{Diffuse gamma-ray intensity $E_\gamma^2 I_\gamma(E_\gamma)$ as a function of the energy (without EBL correction) for the different EGRB components: NG in green, SB in blue, SF-AGN$=$ SF-AGN (SB) + SF-AGN (non-SB) in black, and their sum in magenta. The {\it Fermi} data~\cite{Ackermann:2012vca} are plotted in red.
 \label{fig:diff_contrib_E}}
\end{figure}

To estimate the contribution of the different families of star-forming galaxies as a function of the energy, Fig.~\ref{fig:diff_contrib_E}
shows the total diffuse gamma-ray intensity without EBL attenuation (magenta line) in comparison with contributions from NG (in green), SB (in blue), SF-AGN (in black) where, within our {\it canonical} model, we have assumed that  the SF-AGN component is given by the sum of SF-AGN (SB) and SF-AGN (non-SB) as described in Sec.~\ref{sec:gammaflux}.   
Note as the leading contribution comes from SF-AGN and they dominate the high-energy part of the EGRB together with SB with $\Gamma_{\mathrm{SB}} = 2.2$. 

\section{Diffuse neutrino background from star-forming galaxies}                            
\label{sec:gammanu}
Neutrino production at TeV to PeV energies proceeds via pion production from proton-photon ($p\gamma$) or proton-gas ($pp$)
interactions. Such interactions produce gamma rays as well
\begin{figure}[t]
\centering
\includegraphics[angle=0, width=0.9\textwidth]{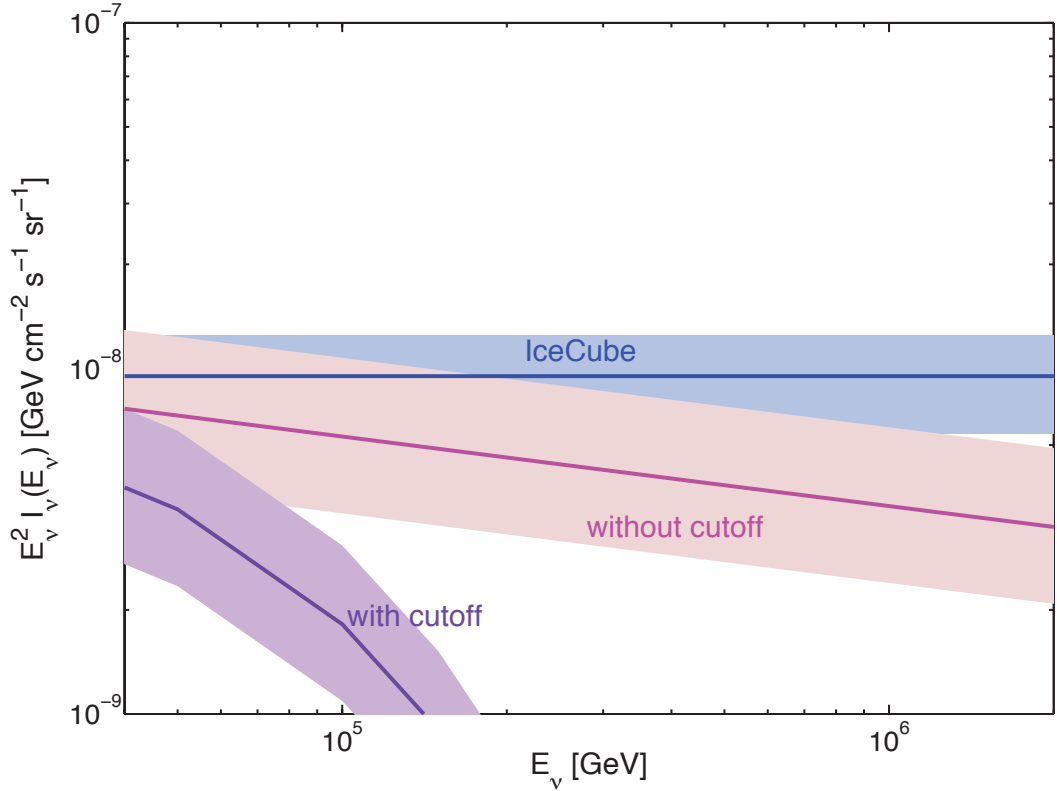}
\caption{Diffuse neutrino intensity $E_\nu^2 I_\nu(E_\nu)$ as a function of the energy.  
The magenta line is the flux obtained adopting the luminosity function approach, the pink band defines the uncertainty band coming from Eq.~(\ref{eq:IRgamma_lum}). The IceCube estimated flux as from~\cite{Aartsen:2014gkd} is marked by the light blue band.  
Our computed flux falls within the astrophysical uncertainties on the IceCube region at $\sim0.5$~PeV energies.  For comparison the diffuse neutrino intensity including an exponential cutoff, $\exp(-E_\nu/80\ {\mathrm{TeV}})$, is plotted in violet.
 \label{fig:neutrino_flux}}
\end{figure}
as neutrinos.
However the relative flux of neutrinos and gamma rays depends on the
ratio of charged to neutral pions, $N_{\pi^\pm}/N_{\pi^0}$, and the
relative neutrino flux per flavor depends on the initial mix of $\pi^+$ to $\pi^-$~\cite{kam+06,Kelner:2006tc}. In this work we focus on $pp$
interactions, as they are the dominant hadronic process for the
star-forming galaxies. In this case, $N_{\pi^\pm} \simeq 2 N_{\pi^0}$ 
and the flavor ratio after oscillations is $\nu_e:\nu_\mu:\nu_\tau=1:1:1$ (for both neutrinos and
antineutrinos)~\cite{kam+06,Kelner:2006tc}.
Hence, following Refs.~\cite{Ahlers:2013xia,Murase:2013rfa,Anchordoqui:2004eu} and ignoring the absorption
during the propagation of photons and neutrinos, the relative differential fluxes of gamma rays and neutrinos are related as
\begin{equation}
\label{nugamma}
\sum_{\alpha} I_{\nu_\alpha}(E_{\nu_\alpha}) \simeq 6\ I_{\gamma, {\rm
no\,abs}}(E_\gamma)\ ,
\end{equation}
with $E_\gamma \simeq 2 E_\nu$, where the subscript ``no abs''
means that we do not include the EBL attenuation in the gamma-ray
intensity.
 
Using Eq.~(\ref{nugamma}), the expected neutrino spectrum due to the star-forming galaxies is
plotted in Fig.~\ref{fig:neutrino_flux} as a function of the energy.
The magenta line is our estimated diffuse neutrino intensity, the pink band defines the uncertainty band coming from
Eq.~(\ref{eq:IRgamma_lum}). The blue line is the IceCube measured flux with its uncertainty band (in light blue) as
in Eq.~(\ref{fitnu}). 
Our estimated diffuse neutrino intensity is comparable, within the astrophysical uncertainties, to the IceCube measurement up to $\sim0.5$~PeV energies. 
However, within our {\it canonical} model assuming $\Gamma_{\rm SB}=2.2$, it is difficult to explain neutrino events around PeV energies.  
Thus, in order to explain all the IceCube data, we need to consider harder spectral indices (see Sec.~\ref{sec:constraints}) or another source component such as galaxy groups/clusters, AGN, and gamma-ray bursts.  
Note that the diffuse neutrino intensity at TeV-PeV energies is dominated by SB and SF-AGN (SB), and contributions from NG and SF-AGN (non-SB) are negligible because of our assumption of $\Gamma_{\rm NG}=2.7$.
In addition, we have assumed that the neutrino spectrum can be extrapolated up to PeV energies.  However, as discussed below (see Sec.~\ref{sec:issues}), this requires that protons are successfully accelerated to $\sim100$~PeV energies, implying that the cosmic-ray spectrum of extragalactic star-forming galaxies is different from that in the MW. For comparison, the diffuse neutrino intensity including an exponential cutoff $\exp(-E_\nu/80\ {\mathrm{TeV}})$ is plotted in violet, motivated by the fact that the MW cosmic-ray nucleon spectrum has a
suppression at the knee~\cite{Kachelriess:2014oma}.

\section{\textit{Fermi} and IceCube bounds on the starburst injection spectra and abundance}                        
\label{sec:constraints}
The injection spectra of the gamma rays of the starbursts are poorly
constrained as well as the fraction of SF-AGN  with an energy spectrum similar to SB
 as discussed in Secs.~\ref{sec:GammaLF} and \ref{sec:highz}.
In this Section, we first treat $\Gamma_{\mathrm{SB}}$ as a free parameter compared to our \emph{canonical} model, 
and then change the fraction of SF-AGN with SB-like injection spectra.  
We compute the resultant diffuse intensities compatible with both the {\it Fermi} and IceCube data.  
It would be interesting to see if one could explain the diffuse neutrino and gamma-ray backgrounds {\it simultaneously}.

\begin{figure}[h!]
\centering
\includegraphics[angle=0, width=0.6\textwidth]{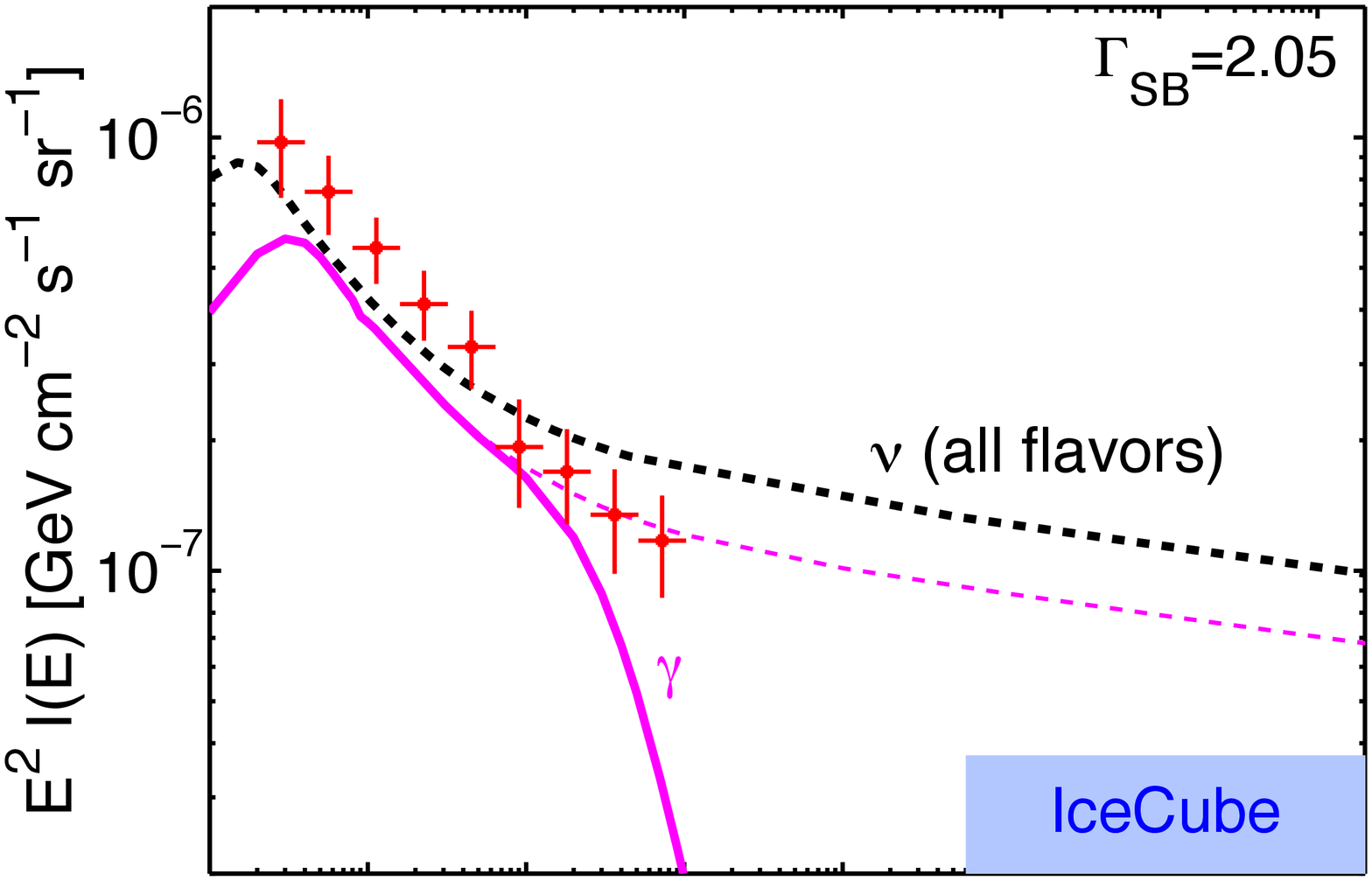}\\
\includegraphics[angle=0, width=0.6\textwidth]{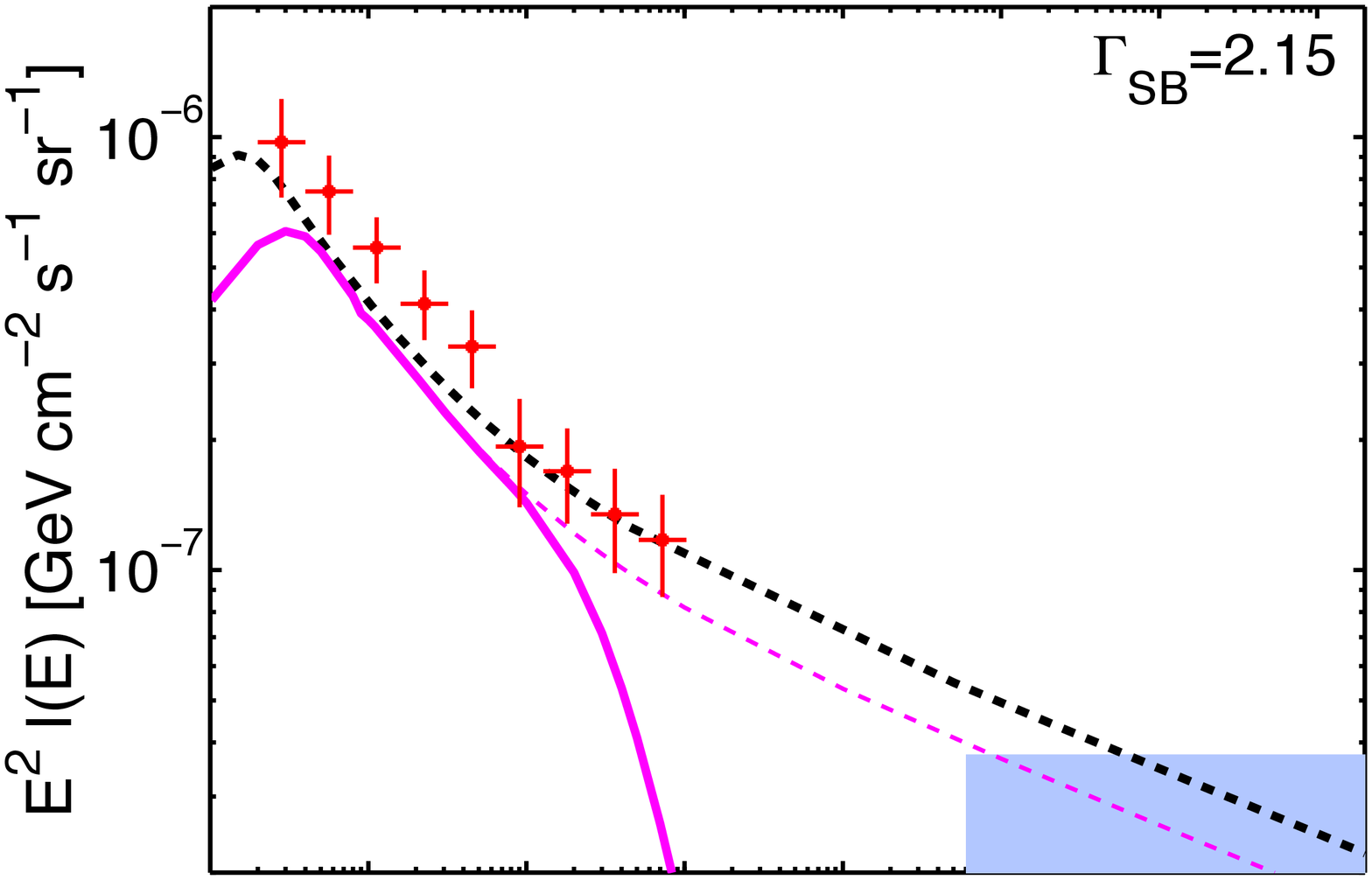}\\
\includegraphics[angle=0, width=0.6\textwidth]{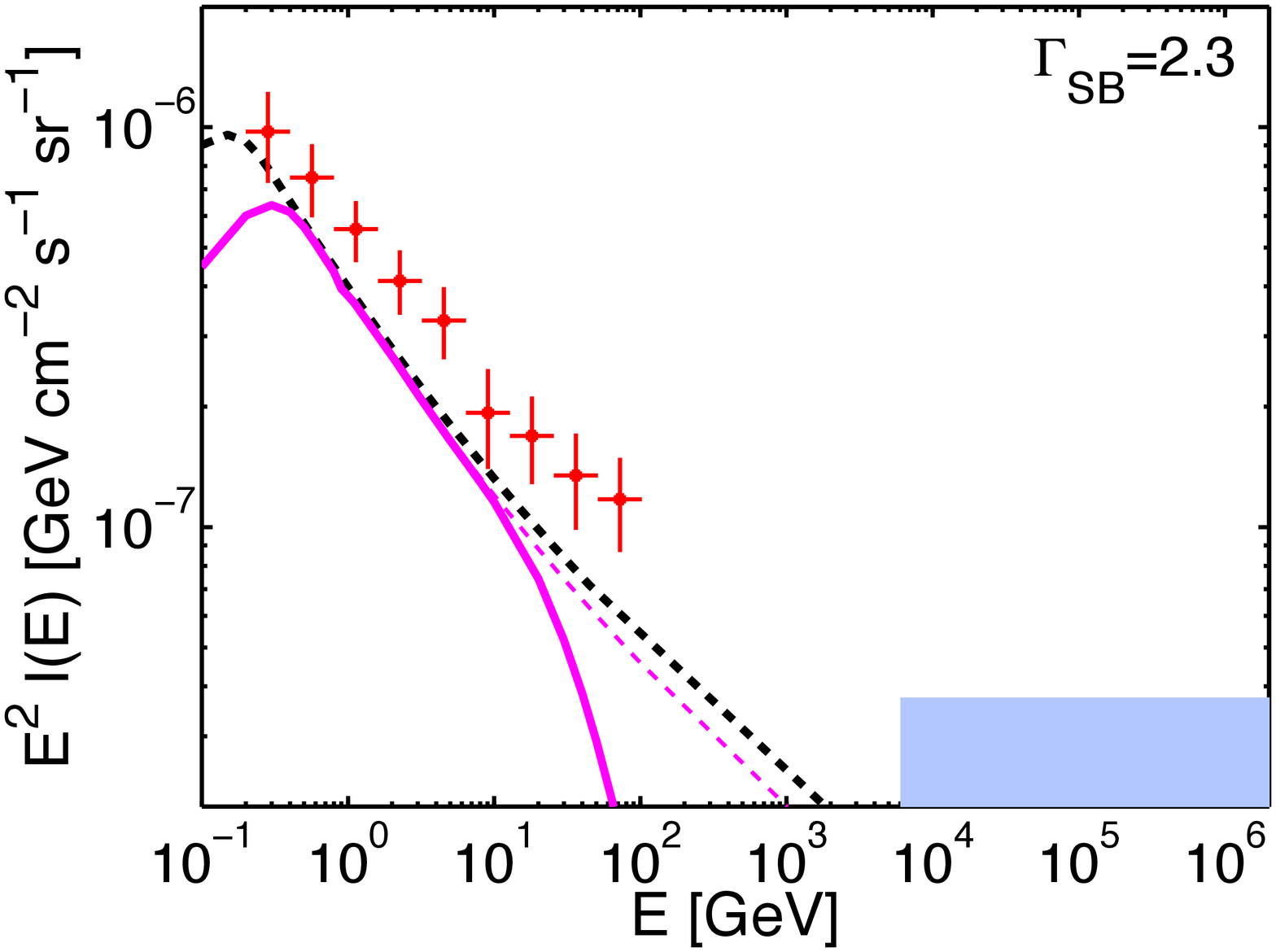}
\caption{
Diffuse gamma-ray (in magenta) and neutrino intensity (in dashed black) $E^2 I(E)$ as a function of the energy for
our {\emph canonical} model, assuming $\Gamma_{\mathrm{SB}} = 2.05, 2.15$ and $2.3$ (from top to bottom). The {\it Fermi} data~\cite{Ackermann:2012vca} are marked in red, while the IceCube region is plotted in light blue~\cite{Aartsen:2014gkd}. 
The EBL attenuation is taken into account for gamma rays (magenta continue lines), the diffuse gamma-ray intensity without EBL attenuation is plotted 
with magenta dashed lines for comparison.
\label{fig:panelsGSB}}
\end{figure}

Figure~\ref{fig:panelsGSB} shows the expected diffuse gamma-ray (in magenta) and neutrino intensity (in dashed black) as
a function of the energy for  our \emph{canonical} model, assuming $\Gamma_{\mathrm{SB}} = 2.05, 2.15$ and $2.3$ (from top to bottom), 
while the {\it Fermi} data~\cite{Ackermann:2012vca} are marked in red and the IceCube region as from Eq.~(\ref{fitnu}) in light blue.    
Although our estimation for the diffuse gamma-ray intensity is always slightly lower than the
{\it Fermi} data, $\Gamma_{\mathrm{SB}} = 2.05$ is currently excluded by the IceCube data (top panel).  
In order to allow such hard spectra, lower ratios of $L_\gamma$ to $L_{\rm IR}$ are needed. 
On the other hand, interestingly, an injection spectral index $\Gamma_{\mathrm{SB}} = 2.15$ can almost explain the {\it Fermi} and IceCube 
data at the same time (middle panel), although some contributions from other gamma-ray source populations are needed to fit the diffuse EGRB spectrum.  
The panel on the bottom shows the diffuse intensities of gamma rays and neutrinos for $\Gamma_{\mathrm{SB}} = 2.3$: The resultant gamma-ray intensity
is lower than the one measured by {\it Fermi} and the corresponding neutrino flux falls below the IceCube band.
In order to give an idea of the role of the EBL attenuation for various spectral indices, in Fig.~\ref{fig:panelsGSB} we plot the diffuse gamma-ray intensity without EBL attenuation (dashed magenta line). Note as it closely follows the diffuse neutrino intensity.

In our \emph{canonical} model we have assumed the spectrum with $\Gamma_\mathrm{SF-AGN(non-SB)} = \Gamma_\mathrm{NG}$ and $\Gamma_\mathrm{SF-AGN(SB)} = \Gamma_\mathrm{SB}$,  as described in Sec.~\ref{sec:gammaflux}.
However, besides $\Gamma_\mathrm{SB}$, also  $\Gamma_\mathrm{SF-AGN}$ is pretty uncertain and might not follow the distribution adopted in our \emph{canonical} model. For example, Seyfert systems (belonging  to the
SF-AGN class) are classified as SF-AGN (non-SB) according to {\it Herschel}~\cite{Gruppioni:2013jna}, while {\it Fermi} classifies the observed Seyferts 
systems NGC
\newpage
\noindent 1068 and NGC 4945 
as starburst galaxies~\cite{Ackermann:2012vca}. Similarly {\it Fermi} finds that the Circinus galaxy, containing an AGN,
has a hard spectral index as SB-like galaxies~\cite{Hayashida:2013wha}. Given the present uncertainties, 
 we now leave the fraction of SF-AGN (SB) as free parameter  constant with 
the redshift as well as its spectral index and discuss the expected EGRB. Therefore, here we assume that  the fraction of galaxies containing AGN that behaves similarly to starbursts is constant with respect to the redshift 
($0 \le f_\mathrm{SF-AGN (SB)} = f_\mathrm{SB-AGN} \le 1$),  and compute their allowed regions compatible with both the {\it Fermi} and IceCube data. 
For this purpose, we rewrite Eq.~(\ref{eq:intensitygamma}) as
\begin{eqnarray}
\label{eq:frac}
I_\gamma(E_\gamma) &=& I_{\gamma,\mathrm{NG}}(E_\gamma) +
 I_{\gamma,\mathrm{SB}}(E_\gamma, \Gamma_{\mathrm{SB}}) +
 \left[f_\mathrm{SB-AGN}\  I_{\gamma,\mathrm{SF-AGN}}(E_\gamma,
  \Gamma_{\mathrm{SB}}) 
 \right.\nonumber\\&&{}\left. +
 (1 - f_\mathrm{SB-AGN})\
  I_{\gamma,\mathrm{SF-AGN}}(E_\gamma,\Gamma_{\mathrm{NG}})\right] \ ,
\end{eqnarray}
where $I_{\gamma, {\rm SF-AGN}} (E_\gamma, \Gamma_{\rm SB})$ adopts the IR luminosity function of SF-AGN but assumes the $E^{-\Gamma_{\rm SB}}$
spectra at high energies.
On the other hand, $I_{\gamma, {\rm SF-AGN}} (E_\gamma,\Gamma_\mathrm{NG})$ is the same except for $E^{-2.7}$ spectra.
\begin{figure}[b]
\centering
\includegraphics[angle=0, width=0.8\textwidth]{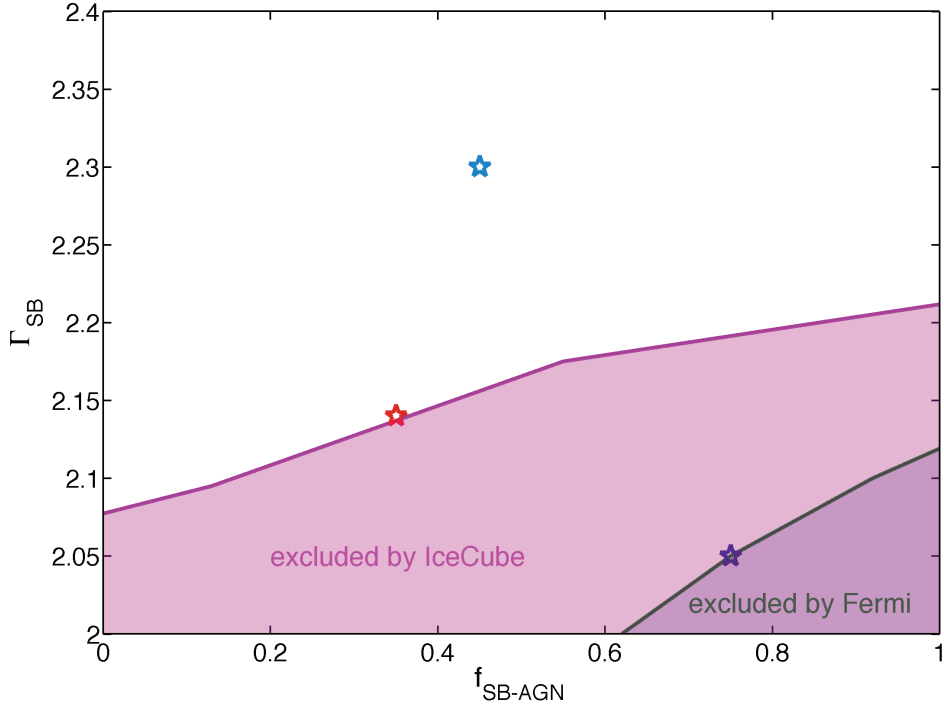}
\caption{Injection spectral index ($\Gamma_{\mathrm{SB}}$) and fraction of SF-AGN behaving similarly to starburst ($f_{\mathrm{SB-AGN}}$) compatible with the {\it Fermi} data~\cite{Ackermann:2012vca} and IceCube ones as in Eq.~(\ref{fitnu}). The region of the parameter space excluded by {\it Fermi}
is marked in green, while the one excluded by IceCube is marked in magenta. Three spots of the scanned parameter space  are selected: $(f_{\mathrm{SB-AGN}},\Gamma_{\mathrm{SB}}) = (0.75,2.05)$, compatible with  {\it Fermi} but excluded by the IceCube data; $(f_{\mathrm{SB-AGN}},\Gamma_{\mathrm{SB}}) = (0.35,2.15)$ that is on the border of the allowed regions of IceCube; $(f_{\mathrm{SB-AGN}},\Gamma_{\mathrm{SB}}) = (0.45,2.3)$ that is below the both data.
\label{fig:exclusion_regions}}
\end{figure}

Figure~\ref{fig:exclusion_regions} shows the parameter regions, in the
plane defined by the injection spectral index ($\Gamma_{\mathrm{SB}}$)
and the fraction of starbursts in the SF-AGN population
($f_{\mathrm{SB-AGN}}$), incompatible with the {\it Fermi}
data~\cite{Ackermann:2012vca} or IceCube data as in Eq.~(\ref{fitnu}).
The {\it Fermi} data tend to exclude a smaller region of the
parameter space than the IceCube ones.  
Hard spectra for SB (i.e., $\Gamma_{\rm SB} \lesssim 2.1$) are excluded from present IceCube data and, as expected, 
both of the data allow harder spectra for smaller starburst fraction, since they give a lower contribution to the high-energy tail of the spectrum.
Note that for $f_{\mathrm{SB-AGN}} = 1$, all SF-AGN behave as SB-like galaxies. Such assumption would be compatible with 
{\it Fermi} data for $\Gamma_{\rm SB} > 2.1$ and with IceCube data for $\Gamma_{\rm SB} > 2.2$. Viceversa, for $f_{\mathrm{SB-AGN}} = 0$, all SF-AGN behave as normal galaxies; this is compatible with 
{\it Fermi} data for any $\Gamma_{\rm SB}$ and with IceCube data for $\Gamma_{\rm SB} > 2.07$.

\begin{figure}[ht!]
\centering
\includegraphics[angle=0, width=0.59\textwidth]{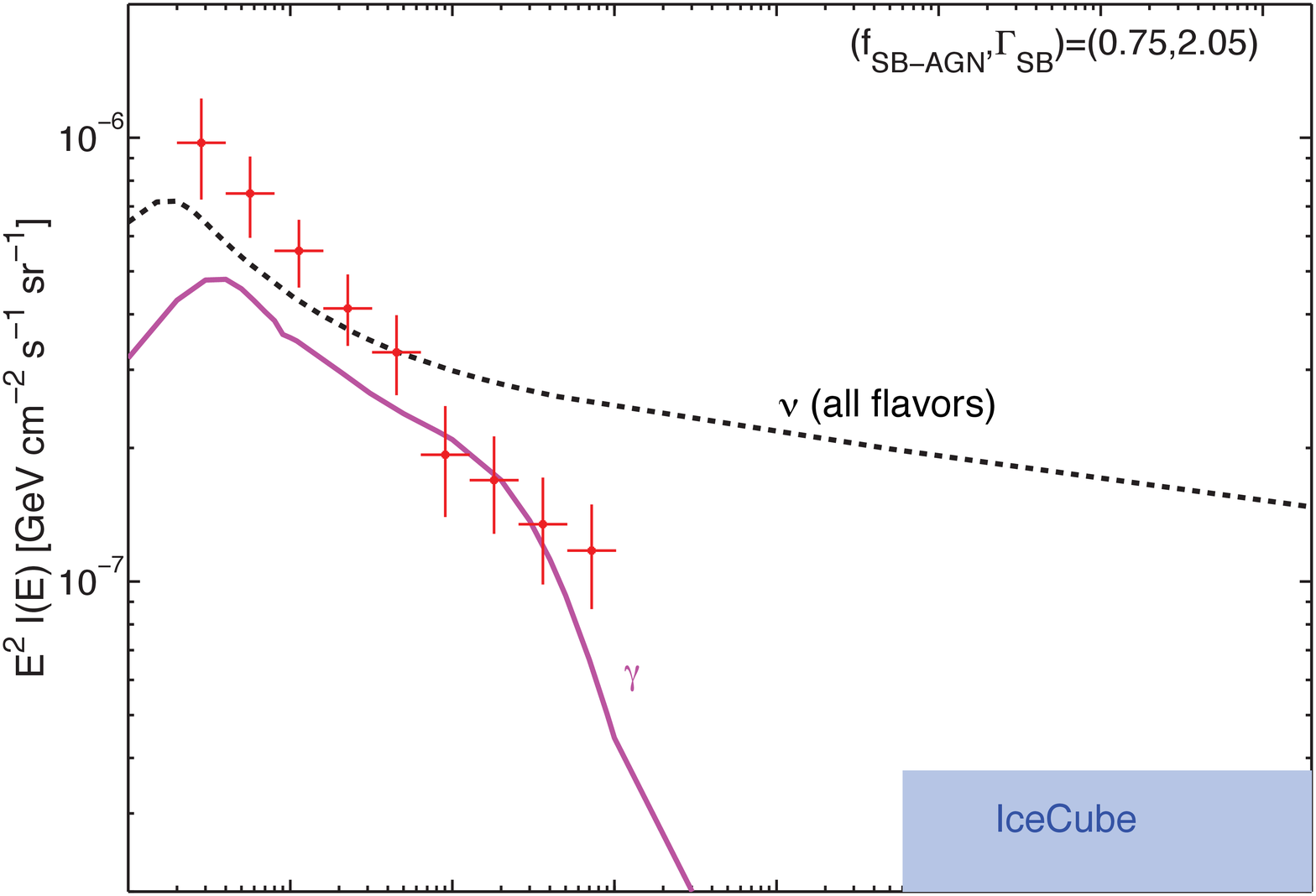}\\
\includegraphics[angle=0, width=0.59\textwidth]{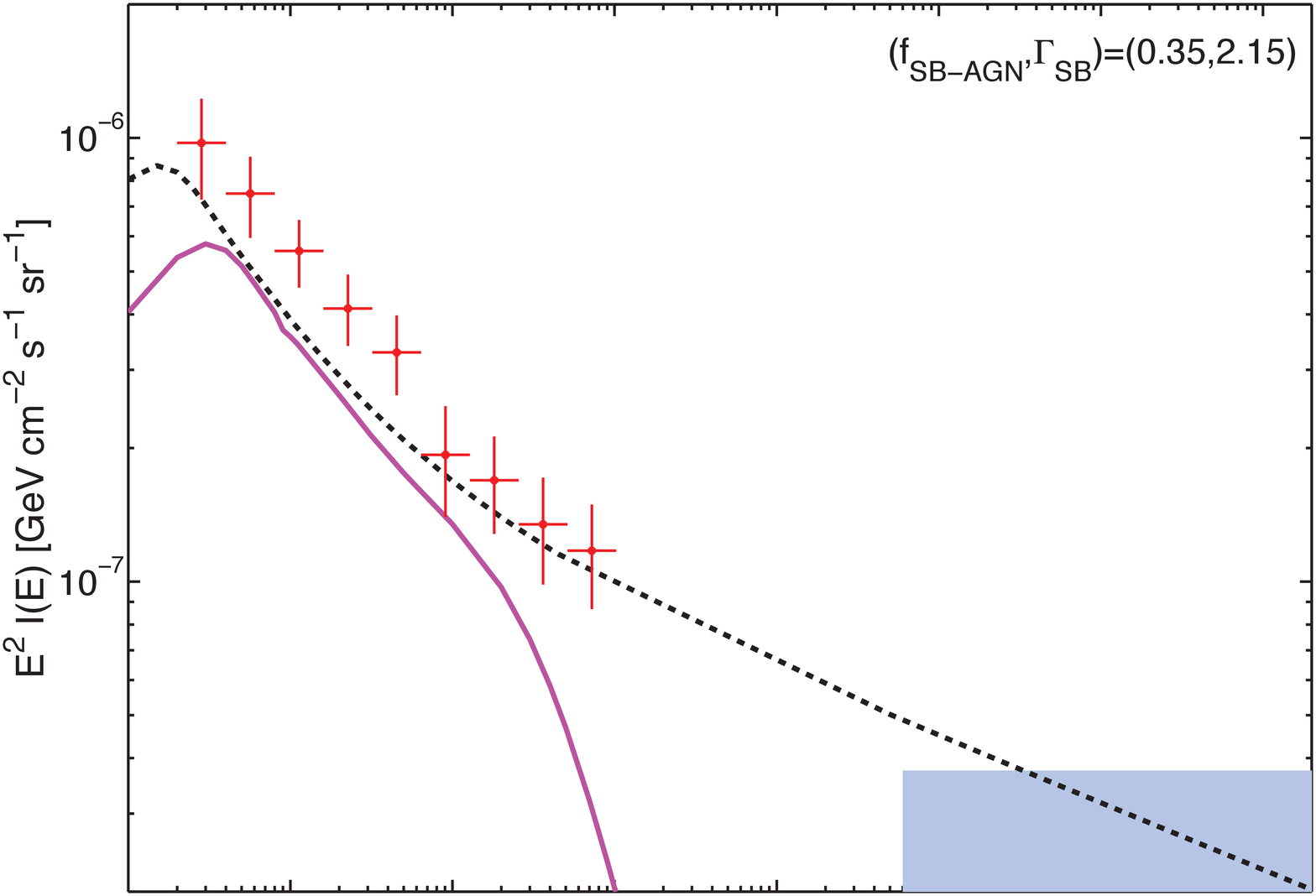}\\
\includegraphics[angle=0, width=0.59\textwidth]{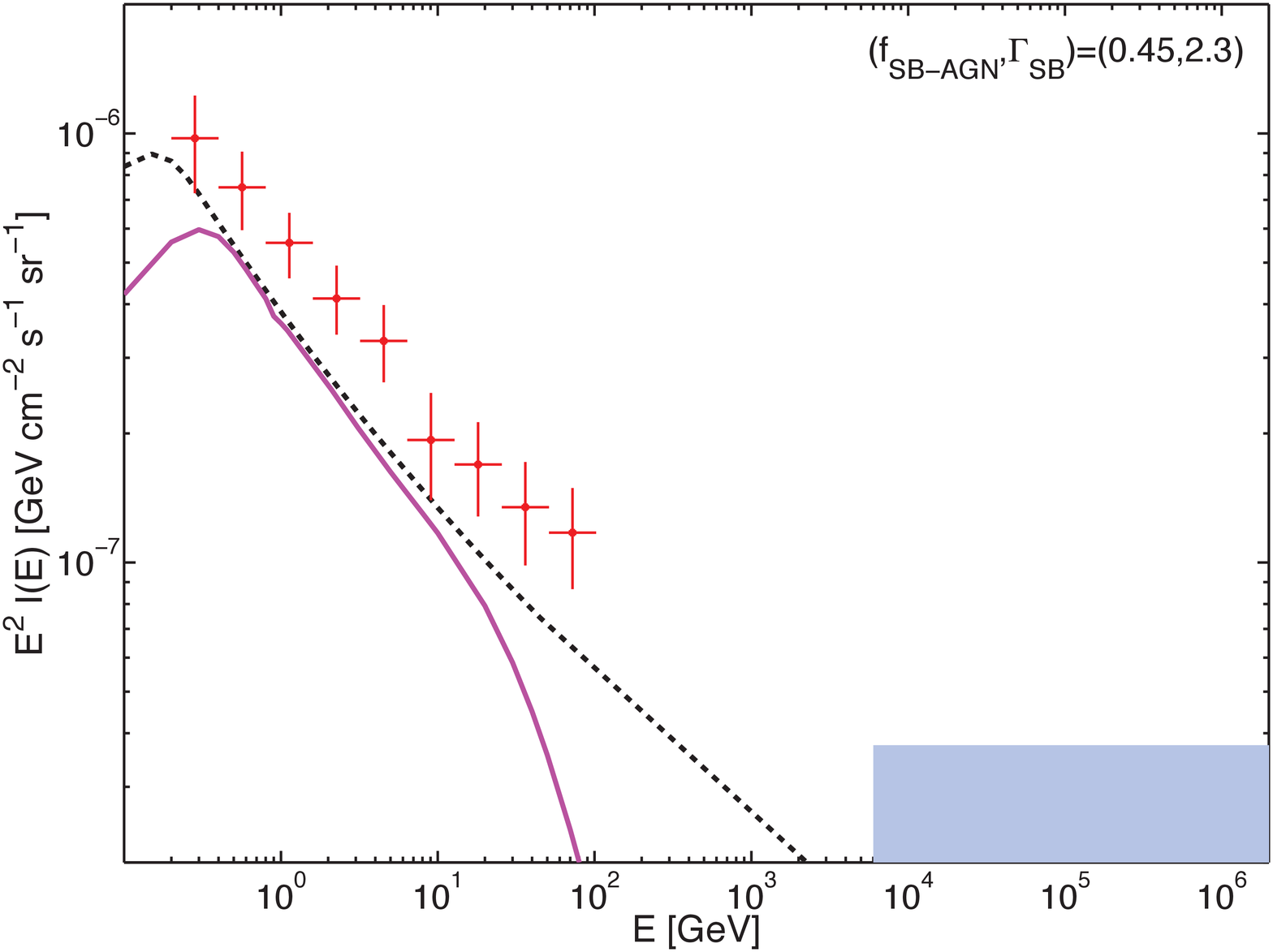}
\caption{Diffuse gamma-ray  (in magenta) and neutrino intensity (in dashed black) $E^2 I(E)$ as a function of the energy assuming a constant fraction of 
SB-AGN as a function of the redshift and various $\Gamma_{\mathrm{SB}}$. From top to bottom,  we assume $(f_{\mathrm{SB-AGN}},\Gamma_{\mathrm{SB}}) = (0.75,2.05), (0.35,2.15)$ and $(0.45,2.3)$, corresponding to the spots marked in Fig.~\ref{fig:exclusion_regions}. The {\it Fermi} data~\cite{Ackermann:2012vca} are marked in red, while the IceCube region is plotted in light blue~\cite{Aartsen:2014gkd}. The EBL attenuation is taken into account for gamma rays.
  \label{fig:panelsfSBfixed}}
\end{figure}

Figure~\ref{fig:panelsfSBfixed} shows the gamma-ray  (in magenta) and neutrino  (in dashed black) intensities for the three spots marked in Fig.~\ref{fig:exclusion_regions}. 
The top panel, $(f_{\mathrm{SB-AGN}},\Gamma_{\mathrm{SB}}) = (0.75,2.05)$
corresponding to the blue spot in Fig.~\ref{fig:exclusion_regions}, shows a diffuse EGRB compatible with the high energy tail of the {\it Fermi} data, 
but a too high diffuse neutrino intensity with respect to the IceCube data.  
The middle panel, $(f_{\mathrm{SB-AGN}},\Gamma_{\mathrm{SB}}) = (0.35,2.15)$
corresponding to the orange spot in Fig.~\ref{fig:exclusion_regions}, shows gamma-ray and neutrino intensities consistent with both 
{\it Fermi} and IceCube data as well as the bottom panel obtained for $(f_{\mathrm{SB-AGN}},\Gamma_{\mathrm{SB}}) = (0.45,2.3)$
corresponding to the light blue spot in Fig.~\ref{fig:exclusion_regions}.  
Also in this case, we confirm the trend found in Fig.~\ref{fig:panelsGSB}:  The injection spectral index matching better the {\it Fermi} and IceCube data simultaneously is $\Gamma_{\mathrm{SB}} \simeq 2.15$.  Note as, for fixed $\Gamma_{\mathrm{SB}}$, Fig.~\ref{fig:panelsGSB} and Fig.~\ref{fig:panelsfSBfixed} look
very similar for the adopted $f_{\mathrm{SB-AGN}}$.

Independently of our work, Ref.~\cite{Murase:2013rfa} obtained constraints of $\Gamma\lesssim2.2$ for general $pp$ scenarios explaining the IceCube data, in the case of star-formation rate evolution.  Interestingly, we have reached the same conclusion in a complementary way.  In our star-forming galaxy modeling, the expected EGRB flux should be lower than the {\it Fermi} data, and the consistency with the IceCube data holds only when $\Gamma_{\rm SB}\lesssim2.2$.  
In particular, the IceCube data can be reproduced by $\Gamma_{\rm SB}=2.15$, which is consistent with the above general constraints on $pp$ scenarios.  The diffuse EGRB data can also be explained by starbursts (including those with AGN), but with some contributions from other populations, and details are affected by EBL uncertainties at high redshifts.  
On the other hand, from Figs.~\ref{fig:panelsGSB}, \ref{fig:exclusion_regions} and \ref{fig:panelsfSBfixed}, only $\Gamma_{\rm SB}\gtrsim2.1$ is allowed whereas smaller spectral indices are ruled out by both IceCube and {\it Fermi} data. 
Thus, in the star-forming scenario, we conclude that $2.1\lesssim\Gamma_{\rm SB}\lesssim 2.2$ is required to explain $\sim100$\% of the IceCube flux by $pp$ interactions in SB and SF-AGN (SB).

\section{High-energy end of the neutrino spectrum}
\label{sec:issues}
As shown above, star-forming galaxies have enough energy budget to explain the diffuse neutrino intensity.  However, there are other
requirements to explain PeV neutrinos.

First, cosmic rays should be accelerated to sufficiently high energies.  Since a significant contribution comes from $z \sim
1$--$2$, the typical neutrino energy can be written as
\begin{equation}
E_\nu\sim 0.04\ E_p \simeq 2~{\rm PeV}~\varepsilon_{p,17}\ [2/(1+\bar{z})]\ ,
\end{equation}
where $\varepsilon_p={10}^{17}~{\rm eV}~\varepsilon_{p,17}$ is the
proton energy in the cosmic rest frame and $\bar{z}$ is the typical
source redshift.  Thus, to have $\sim2$~PeV neutrinos observed by
IceCube, the energy per nucleon should exceed the {\it knee} energy at
$3\mbox{--}4$~PeV.  In the MW, it has been believed that Galactic
supernova remnants are the main origin of Galactic cosmic
rays~\cite{Bluemer:2009zf}, but the above equation means that galaxies
responsible for PeV neutrinos have to contain more powerful
accelerators than in the MW.  Rather, neutrinos around a possible
$\sim1\mbox{--}2$~PeV break may be related to the origin of cosmic
rays close to the {\it second knee}~\cite{Murase:2013rfa}.

Second, cosmic rays around $\sim10$--$100$~PeV should be confined in galaxies, so
that they can be sources of PeV neutrinos with sufficiently hard
spectra. This implies that magnetic-field properties in these
extragalactic starbursts should be different from that in the MW.  
 The Galactic cosmic-ray composition is heavier above the {\it knee}, so
the cosmic-ray nucleon
\newpage
\noindent  spectrum is significantly steeper above that
energy.  Since the neutrino energy is less for nuclei with the same
energy, adopting the observed MW cosmic-ray spectrum leads to a break
around 80~TeV in the neutrino spectrum, which looks insufficient to
explain the IceCube data without introducing another component (see Fig.~\ref{fig:neutrino_flux}).

Below, we describe these issues in more detail.  Some of them are
pointed in Ref.~\cite{Murase:2013rfa}, but more extended discussions
are useful for the completeness of our work since the extrapolation
from GeV to PeV energies is not guaranteed.  In particular, we here
point out that obscured or low-luminosity AGN in star-forming galaxies
may supply sufficiently high-energy cosmic rays.

\subsection{Maximum proton energy}
\label{sec:maximum}
Can we expect the proton maximum energy $\varepsilon_p^{\rm
max}\sim10$--$100$~PeV?  Ref.~\cite{Murase:2013rfa} discussed the
possibility that peculiar supernovae are responsible for
$\sim10$--$100$~PeV cosmic rays.  For example, gamma-ray bursts and
transrelativisitic supernovae (or low-luminosity gamma-ray bursts) may
accelerate cosmic rays up to ultrahigh energies, and their cosmic-ray
energy generation rate can be $\sim1$--$10$\% of that of core-collapse
supernovae~\cite{mur+08,wan+07}
relevant for the high-energy cosmic-ray spectra beyond the {\it
knee}~\cite{sve03}.
In particular, broadline Type Ibc, often called hypernovae (that are
not necessarily relativistic), have kinetic energy
$\sim{10}^{52}$~erg~\cite{maz+07,fan+11} and their rate is 1--2\% of
core-collapse supernova rate~\cite{smi+10}.  At the radius where the
ejecta starts to decelerate, the so-called Sedov radius $R_{\rm
Sed}$~\cite{Taylor1950,Sedov1946}, the proton maximum energy is
estimated to be $\varepsilon_p^{\rm max}\approx (3/20) (V_{\rm ej}/c)
e B R_{\rm Sed} \simeq 9.7~{\rm PeV}~B_{-3.5} {\mathcal E}_{\rm
ej,52}^{1/3} V_{\rm ej,9.5}^{1/3} n^{-1/3}$, which is the order of
$\sim10$~PeV.  Here ${\mathcal E}_{\rm ej}=10^{52}~{\rm erg}~{\mathcal
E}_{\rm ej,52}$ and $V_{\rm ej}=10^{9.5}~{\rm cm}~{\rm s}^{-1}~V_{\rm
ej,9.5}$ are the ejecta energy and velocity, and $n$ is the ambient
density~\cite{sve03}.
However, these peculiar supernovae are much rarer than normal
supernovae.  Even though they could be more common at higher
redshifts, it is not clear if the energy budget of these peculiar
supernovae is comparable to that of normal supernovae.  Alternatively,
other speculations, including higher values of
$B\sim1\mbox{--}30$~mG~\cite{Murase:2013rfa} and superbubbles
consisting of multiple supernovae~\cite{bf92}, have been discussed.

In this work, we have demonstrated that SF-AGN may give an important
contribution to the gamma-ray and neutrino backgrounds.  
They are needed especially if $\Gamma_{\rm SB}\gtrsim2.1$ (see Fig.~\ref{fig:exclusion_regions}).  
AGN are also potential cosmic-ray accelerators, as has been motivated by recent
gamma-ray observations of nearby Seyfert galaxies, NGC 1068 and 4945
(e.g.,~\cite{yoa+14} and references therein).  Although the fraction
of AGN associated with starbursts is highly uncertain, they are
potentially important~\cite{Murase:2014foa}, and we here give
quantitative estimates.
Most of AGN in star-forming galaxies are radio-quiet quasars or
Seyfert galaxies, which are obscured or low-luminosity AGN.  Such
radio-quiet AGN may have weak jets with luminosity
$L_j\sim{10}^{42}$--${10}^{\rm 43}~{\rm erg}~{\rm s}^{-1}$
(e.g., Ref.~\cite{mp14} and references therein), and very-high-energy
cosmic rays may be accelerated at shocks caused by the
jets~\cite{pee+09}.  Assuming the jet opening angle $\theta_j\sim0.1$
and dissipation radius $R$, the magnetic field is written as
$B_j=(4\epsilon_B L_j/(\theta_j^2 R^2 c))^{1/2}\simeq0.12~{\rm
mG}~\epsilon_{B,-2}^{1/2}L_{j,43}^{1/2}\theta_{j,-1}^{-1}{(R/100~{\rm
pc})}^{-1}$, where $\epsilon_B$ is the magnetic energy fraction.
Applying the formula of the non-relativistic shock to the relativistic
shock (which gives a rough estimate), we have
\begin{equation}
\varepsilon_p^{\rm max}\approx(3/20)eB_jR\simeq1.6~{\rm
EeV}~\epsilon_{B,-2}^{1/2}L_{j,43}^{1/2}\theta_{j,-1}^{-1},
\end{equation}
so that $\gtrsim100$~PeV protons may be accelerated by the weak jets.
Note that such a low-power jet forms a bubble inside the intergalactic
medium, and it will take many years to penetrate the host galaxy.
In addition, recent observations have shown that fast outflows from
AGN are ubiquitous~\cite{tom+13,tom+14,har+14}.  The kinetic
luminosity of such presumably disk-driven winds ($L_w$) is
$\sim1$--$10$ percent of the AGN radiation luminosity, and cosmic rays
may potentially be accelerated.  At the dissipation radius
$R\sim100$~pc, the outflow velocity is of the order of $V_w\sim1000~{\rm
km}~{\rm s}^{-1}$~\cite{tom+13}.  This leads to a magnetic field
$B_w=(2\epsilon_B L_w/(R^2 V_w))^{1/2}\simeq0.46~{\rm
mG}~\epsilon_{B,-2}^{1/2}L_{w,44}^{1/2}{(R/100~{\rm
pc})}^{-1}{(V_w/1000~{\rm km}~{\rm s}^{-1})}^{-1/2}$.  Then, assuming
the first-order Fermi acceleration mechanism at shocks, the maximum
energy is estimated to be
\begin{equation}
\varepsilon_p^{\rm max}\approx(3/20)(V_{w}/c)eB_wR\simeq21~{\rm
PeV}~\epsilon_{B,-2}^{1/2}L_{w,44}^{1/2}{(V_w/1000~{\rm km}~{\rm
s}^{-1})}^{1/2}.
\end{equation}
In both the jet and wind cases, $\sim10$--$100$~PeV protons may be
produced, where SF-AGN could be sources of PeV neutrinos.  Note that a
lot of possibilities have been suggested but they are not mutually
exclusive.  Possibly, if we consider transients, one source class
could be the origin of cosmic rays from GeV to ultrahigh
energies~\cite{kat+13}. Our results are useful even in this case.

\subsection{Cosmic-ray confinement}
\label{sec:diffusion}
Even if the maximum proton energy somehow achieves $\sim100$~PeV at
the sources, there remains a theoretical question whether it is
possible to trap $\sim100$~PeV cosmic rays in the galaxies. The
criterion is basically determined by comparing the $pp$ cooling time
($t_{pp}$) and cosmic-ray escape time ($t_{\rm esc}$).  The pionic
loss time is given by $t_{pp} \approx 2.7~{\rm
Myr}~\Sigma_{g,-1}^{-1}~(h/{\rm kpc})$, where $\Sigma_g$ is the column
density and $h$ is the scale height.  The escape time depends on
properties of magnetic fields.   If cosmic rays are well-trapped in
the fluid, their escape is governed by advection losses via
starburst-driven or AGN-driven outflows.  Using the wind velocity
$V_w$, the advection escape time is estimated to be $t_{\rm adv}
\approx h/V_w \simeq 0.98~{\rm Myr}~(h/{\rm kpc})~V_{w,8}^{-1}$.

However, especially at high energies, the diffusion escape is expected
to be more important.  Here, we rely on the analogy with the diffusion
model for the MW.  To be consistent with $\Gamma_{\rm SB}=2.2$ and 
$\Gamma_{\rm NG}=2.7$ in our canonical model, let us consider 
$\delta_{\rm CR} = 0.5$ as an example (see Ref.~\cite{Murase:2013rfa} for the Kolmogorov case).  The
confinement of 100~PeV protons requires the critical energy of
$\varepsilon_c = e B l_{\rm coh} > 100~{\rm PeV}$, leading to the
coherence length $l_{\rm coh} \gtrsim 0.34~{\rm pc}~B_{-3.5}^{-1}
\varepsilon_{p,17}$.  The diffusion coefficient at $\varepsilon_c$ is
$D_c = (1/3) l_{\rm coh} c$, so we have $D =
D_c{(\varepsilon_p/\varepsilon_c)}^{\delta_{\rm CR}}$ for
$\varepsilon_p < \varepsilon_c$. For the MW, the diffusion coefficient
at GeV is $D_0 \sim {10}^{28}~{\rm cm}^2~{\rm s}^{-1}$.  Magnetic
fields of starbursts can be $\sim 100$ times higher, where one may
naively expect $D_0 \sim {10}^{27}~{\rm cm}^2~{\rm s}^{-1}$.  But it
turns out to be insufficient, since this value is too large for cosmic
rays to get confined.  In principle, $D_0$ can be as low as $D_0
\gtrsim {10}^{24}~{\rm cm}^2~{\rm s}^{-1}~B_{-3.5}^{-1/2}$.  Thus,
assuming $\delta_{\rm CR}=0.5$ and $D_0 \sim {10}^{25}~{\rm cm}^2~{\rm
s}^{-1}$ optimistically, the diffusion escape time is estimated to be
\begin{equation}
t_{\rm diff} \approx \frac{h^2}{4D} \simeq 0.75~{\rm
Myr}~D_{0,25}^{-1}~\varepsilon_{p,17}^{-1/2}{\left(\frac{h}{\rm
kpc}\right)}^2\ .
\end{equation}
Then, for proton spectra, one would expect a spectral break at
$\varepsilon_p^b \sim 7.7~{\rm PeV}~D_{0,25}^{-2}
\Sigma_{g,-1}^2{(h/{\rm kpc})}^2$ if $t_{pp} < t_{\rm adv}$. If
$t_{\rm adv} < t_{pp}$, we have $\varepsilon_p^b \sim 59~{\rm
PeV}~D_{0,25}^{-2} V_{w,8}^{2}{(h/{\rm kpc})}^2$.

Prototypical starbursts such as M82 and NGC 253 indicate a column
density of $\Sigma_g \sim 0.1~{\rm g}~{\rm cm}^{-2}$ and a scale
height of $h \sim 50$~pc, so they have difficulty in trapping
$\sim100$~PeV cosmic-ray protons.  Instead, the confinement of
$\sim100$~PeV cosmic rays may be possible in high redshift starbursts,
if the diffusion coefficient around $\sim100$~PeV is sufficiently
close to that in the Bohm limit~\cite{Kulsrud}.
Although details are highly uncertain and beyond the scope of this
work, such a situation could be realized, if turbulence is caused by
merger shocks or outflows from AGN or continuously generated by
supernova shocks.  We hope that our work encourages further studies to
give physical understandings of plasma and magnetohydrodinamic
properties in starburst galaxies.

Finally, we summarize the high-energy end of neutrino spectra.  For
$\varepsilon_p^{\rm max}\ll\varepsilon_p^{b}$, we roughly expect
\begin{equation}
E_\nu^2 I_\nu\propto E_\nu^{-\Gamma_{\rm SB}} \exp(-E_\nu/E_\nu^{\rm max}),
\end{equation}
where $E_\nu^{\rm max}\sim0.05\varepsilon_p^{\rm max}/(1+\bar{z})$.
For $\varepsilon_p^{\rm max}\gg\varepsilon_p^{b}$, a broken power-law
spectrum (that is smoothened by the line-of-sight integral)
\begin{equation}
E_\nu^2 I_\nu\propto
\left\{\begin{array}{ll}
{(E_{\nu}/{E}_{\nu}^{b})}^{-\Gamma_{\rm SB}}
& \mbox{($E_\nu \le {E}_{\nu}^{b}$)}
\\
{(E_{\nu}/{E}_{\nu}^{b})}^{-\Gamma_{\rm SB}-\delta_{\rm CR}}
& \mbox{(${E}_{\nu}^{b} < E_\nu$)}
\end{array} \right.
\end{equation}
is expected, where $E_\nu^{b}\sim0.05\varepsilon_p^{b}/(1+\bar{z})$. 
Interestingly, the present IceCube data indicate a possible break or cutoff energy 
at a few PeV~\cite{Aartsen:2014gkd}, for sufficiently small spectral indices 
$\Gamma\lesssim2.2$.  
If this is the case, revealing this spectral structure is important, and
future neutrino observations by IceCube or next-generation detectors
will be able to discriminate between the two possibilities.

\section{Conclusions}
\label{sec:conclusions}
The importance of the EGRB is related to the fact that it encodes unique information about the high-energy processes in the universe, 
it has an isotropic sky distribution and is dominated by gamma rays induced by cosmic-ray protons and electrons interacting with the target gas and radiation fields.
Although main contributions of the diffuse EGRB are likely to come from star-forming galaxies and blazars, such a diffuse background should also include other sub-leading contributions from unresolved extragalactic sources in the different energy range, such as radio galaxies, gamma-ray bursts, intergalactic shocks produced by structure formation, and dark-matter annihilation.

Star-forming galaxies are expected to largely contribute to the $0.3$--$100$~GeV gamma-ray background, because of interactions of cosmic-rays with the diffuse gas.  Besides normal galaxies, a special subset of star-forming galaxies are the less numerous, but individually more luminous, starburst galaxies. 
The starburst galaxies undergo an epoch of star formation in a very localized region at an enhanced rate in comparison to the normal galaxies, due to galaxy mergers or bar instabilities.  According to the classification done by {\it Herschel} in terms of the luminosity functions, the largest fraction of star-forming galaxies is made by galaxies containing low luminosity or dim AGN, and such family could have an energy spectrum similar to normal galaxy or to starburst galaxies according to the cases.  Our understanding of galaxy formation has dramatically improved in the last few years. 
In particular, the {\it Herschel} PEP/HerMES survey has recently provided an estimate of the IR luminosity function up to $z \simeq 4$, finally allowing to explore
the galaxy composition for $z \ge 2$, suggesting that galaxies containing AGN are largely abundant and that the starburst fraction can reach up to $30\%$ of the total at high $z$. 

We have provided an estimation of the EGRB adopting IR data at $z \ge 2$ through the luminosity function provided in~Ref.~\cite{Gruppioni:2013jna}.  The luminosity function indeed provides a powerful tool to probe the distribution of galaxies over cosmological time, since it allows to access the statistical nature of galaxy formation and their evolution.  Our estimation is compatible with the {\it Fermi} data in the 0.3--30~GeV range including the astrophysical uncertainties.  
In our canonical model, the sum of NG and SB components leads to $\Gamma_{\rm SF}\sim2.4$, where we find that high-energy data are not explained due to the EBL attenuation.  We identify the very-high-energy gamma-ray excess, which may indicate another population like BL Lac objects, although quantitative details are affected by EBL uncertainties.    

High-energy neutrinos are also expected to be emitted by these sources together with gamma rays.  Assuming that at least $\sim100$~PeV cosmic rays are accelerated and confined in starburst galaxies,  we estimated the expected diffuse neutrino background from star-forming galaxies, and find that it  can be in agreement with the IceCube data, especially at low energies.  
Although such estimation needs to be taken with caution since better statistics of the IceCube high-energy events is required, the disagreement at high energies might suggest that the observed IceCube flux is given by the contribution of other sources such as gamma-ray bursts and AGN.  In particular, in our star-forming galaxy scenario, we found that for $\Gamma_{\mathrm{SB}} \simeq 2.15$ both {\it Fermi} and IceCube data are consistently explained. 

The {\it Fermi} and IceCube data could constrain the abundance and the injection spectral index of the fraction of galaxies containing AGN that
behave similarly to starbursts.  
We find that hard spectra for SB and SB-like SF-AGN (i.e., $\Gamma_{\rm SB}\lesssim2.1$) are excluded from present IceCube data and, as expected, the data allow more abundant SB-like galaxies for softer spectra, since they give a lower contribution to the high energy tail of the spectrum.

In conclusion, our understanding of how galaxies evolve has dramatically increased over the last decade and, after the new discovered IceCube high-energy neutrinos, we have shown as a powerful multi-messenger connection between gamma rays and neutrinos can provide important new ways to constrain star-forming galaxy distribution as well as their spectral parameters.

\section*{Acknowledgments} 
We thank John Beacom, Caitlin Casey and Nick Scoville for useful discussions,
and the anonymous referee for insightful comments.
This work was supported by
the Netherlands Organization for Scientific Research (NWO) through a
Vidi grant (I.T. and S.A.) and by NASA through Hubble Fellowship Grant
No. 51310.01 awarded by the STScI, which is operated by the Association
of Universities for Research in Astronomy, Inc., for NASA, under
Contract No. NAS 5-26555 (K.M.).


\end{document}